\documentclass[twocolumn]{aa} 

\usepackage{graphicx}
\usepackage{txfonts}
\usepackage{threeparttable}
\begin{document}

   \title{Close binary evolution based on Gaia DR2:}

   \subtitle{the origin of late WC-type Wolf-Rayet stars with low luminosity}
   \author{
   Weiguo Peng
          \inst{1}
          \and
          Hanfeng Song\inst{1,\dag}
           \and
Georges Meynet\inst{2,\dag}
 \and
Andre Maeder\inst{2}
 \and
Fabio Barblan\inst{2}
 \and
Ruiyu Zhang\inst{3}
 \and
Sylvia Ekstr\"omt\inst{2}
 \and
Cyril Georgy\inst{2}
 \and
Gang Long\inst{1}
 \and
Liuyan Zhao\inst{1}
 \and
Ying Qin\inst{4}
          }
   \institute{College of Physics, Guizhou University, Guiyang city, Guizhou Province, 550025, P.R. China
         \and
             Geneva Observatory, Geneva University, CH-1290 Sauverny, Switzerland
             \and
         Shanghai Astronomical Observatory, Chinese Academy of Sciences, Shanghai city, 200030, P.R. China
          \and
Department of Physics, Anhui Normal University, Wuhu city, Anhui Province, 241000, P.R. China\\
$^{\dag}$Corresponding author,
\email{hfsong@gzu.edu.cn; georges.meynet@unige.ch}
             }

   \date{Received, accepted }


 \abstract
{
The observed late-type WC Wolf-Rayet stars (WC7-9) with low luminosity below $\rm \log L/L_{\odot} < 5.4$ in the HR diagram cannot be reproduced satisfactorily by the evolutionary track of single stars. The mass transfer due to Roche lobe overflow drastically modifies the internal structure and surface compositions of two components. Therefore, binaries provide a very promising evolutionary channel to produce these WC stars.
}
 {
 The Gaia satellite provides accurate distances to WC stars and confirms the luminosities of WC stars. Based on a small grid containing single stars and binaries, we aim to investigate the extent to which the evolution of a single or a close binary can reproduce the properties of these stars.
}
 {
We considered single-star models with masses between 20 and 40 $M_{\odot}$. We calculated the evolution for three binaries with a 30 $M_{\odot}$ primary star with a 27 $M_{\odot}$ companion star with initial orbital periods of 6.0, 20.0, 500.0, and 1000.0 days.
 }
{
The rotating single star can evolve into a late-type WC star but with high luminosity (i.e., $\rm \log L/L_{\odot} >5.4$). Enhanced wind mass loss rates during RSG and WR stages, as proposed in the literature, can cause the star to approach the observational range of low-luminosity WC stars and favor the formation of low-luminosity WO stars.
In a wide binary system with initial $\rm P_{orb}=1000$ days, the primary star can evolve into a late-type WC star and be compatible with the observed properties of the low-luminosity WC stars. The result is almost insensitive to the adopted accretion efficiency $1-\beta$.
}
{
Compared with single stars, the low brightness is due to a smaller temperature gradient inside the star after the Case C Roche lobe overflow, while the low effective temperature is due to envelope expansion.
There are four physical reasons for the formation of the expanding envelope. Firstly, less helium envelope can be transferred to the companion star in this system. Heavy helium envelopes can be heated by the helium burning shell and this creates the necessary conditions for the envelope expansion. Secondly, the expansion of the helium envelope can also be boosted by the sharp shrinkage of the larger carbon-oxygen core through the mirror effect. Thirdly, a more massive WC star can attain a higher Eddington factor because of its higher L/M ratio. The increase in $\rm L/M$ with mass is the primary cause for the extended envelopes in WC stars. Finally, the iron opacity bump at $\rm T \sim 10^{5.25}$ K may also trigger envelope inflation because it can lead to a larger Eddington factor.
 }


\maketitle
%

\section{Introduction}
Wolf-Rayet (WR) stars are a small group of stars with high luminosity and surface temperatures that can reach 90 000 K. They have strong and broad emission lines of ionized helium, carbon, oxygen, and nitrogen but few absorption lines. Based on the presence of
prominent emission lines from nitrogen or carbon, WR stars are divided into the nitrogen (WN), carbon (WC), and oxygen (WO)
sequence with subclasses defined by the ratios of characteristic
emission lines. While some WN stars have considerable amounts
of hydrogen, all WC and WO stars do not display any measurable
amount of hydrogen in their spectra. High-ionization WC4-6 subtypes are defined as early-type WC stars, and low-ionization WC7-9 subtypes are late-type WC stars. The broad emission lines are produced in optically thick stellar winds with very high speeds of up to 3000 $\rm  km s^{-1}$, which means the star is continuously and rapidly losing mass.

WR stars are stripped stellar cores and are produced by strong stellar winds or binary mass transfer. WC stars are a WR subtype with strong carbon spectral line features. Helium burning products are present on these objects, which is a characteristic of evolved massive stars.
In particular, a challenging problem for the evolution of massive stars is the existence of low-luminosity WC stars in the region $4.6 <\log T_{\rm eff}< 5.1$ with $4.9 <\log L/L_{\odot} \leq 5.4$ (Sander et al. 2012; Yoon 2017a). Different origins has been proposed in the literature to explain the existence of low-luminosity (and thus low-mass) WC stars. From a theoretical point of view, we can summarize these origins as follows:

-The red supergiant (RSG) scenario: The mass loss by the typical prescription used during the
 RSG wind formalism proposed by de Jager et al. (1988) may be underestimated. This formalism predicts the real
 RSG wind-loss rates within no better than a factor of between five and ten. As shown by Meynet et al. (2015), the increase in the mass-loss rate should be more than ten times higher during RSG phase with respect to the standard one. These authors found that enhanced mass-loss rates during the RSG phase have little impact on the Wolf-Reyet populations.  The low-luminosity WC stars cannot be reproduced by current rotating single-star models, which merely give rise to high-luminosity WC stars with luminosities exceeding $\rm 10^{5.4} L_{\odot}$ (Meynet \& Maeder 2005; Georgy et al. 2012).

-The massive star scenario: Stars with masses above $25 M_{\odot}$ show greater mass loss. Meynet et al. (1994) showed that low-luminosity WC stars with $\log L/L_{\odot}$ down to 4.5 might be produced from progenitors of 120 $M_{\odot}$ at solar metallicity by using stronger mass-loss rates. However, the high mass-loss rates adopted by Meynet et al. (1994) might no longer be supported by the more recent mass-loss determinations for O-type and WR stars that account for the effect of clumping (Vink et al. 2001; Nugis \& Lamers 2000). It may be the case that the mass-loss prescriptions used underestimate the true time-averaged mass-loss rate due to the fact that they ignore mass-loss episodes of short duration. Very massive stars could lose large amounts of mass in very short periods during strong outbursts at the stage of luminous blue variable (LBV; Smith \& Owocki 2006). LBVs might help to remove the necessary amount of mass as they occur over relative wide range of luminosities (Humphreys et al. 2016; Smith et al. 2019). However, the LBV mass-loss rate is not yet well understood and we are not in a position to resolve this problem.

-The close-binary scenario: The low-mass WC stars are produced by close binary systems through Roche lobe overflow (RLOF). Close binary stars are important in understanding the formation, evolution and death of massive stars.Vanbeveren and Conti (1980) considered the Galactic census of WR binaries, concluding that the real Galactic WR+OB frequency is no more than $30\%-40\%$, a percentage that still holds today. The investigation of binaries in the population of O-type and WR-type stars began in early sixties (Dalton and Sarazin 1995; Vanbeveren 1995; Vanbeveren et al. 1997). Vanbeveren et al. (1998a, b) found that more than $70 \%$ (resp. $60 \%$)of O stars (resp. early B-type stars) are born with a companion star. These authors investigated the O-type binary frequency needed to meet the WR+O-type binary frequency in the solar neighborhood. Eldridge et al. (2008) found that the minimum initial mass for stars to become WC stars is lowered from about 27 solar masses in the single-star scenario to 15 solar masses in the close-binary scenario. This gives a milder appreciation of the relative importance of the different effects for forming WR stars.

Because of its significant incidence, binarity has to be taken into account when deriving properties from massive star populations (de Mink er al. 2014). It is undeniable that understanding binarity and its effects could help to explain the different observed characteristics in massive star populations which remain poorly understood from studies of the single-star scenario.
Binarity may help to explain the formation of WR stars in situations where the single-star scenarios have difficulty in this respect. Stellar winds are weaker at low metallicity than those at solar metallicity. Moreover, star with lower initial masses cannot lose enough hydrogen envelope to uncover their central cores in the absence of mass transfer via RLOF.
Due to interactions in close binary systems, the internal structures of the
stars, including the sizes of their cores, can be significantly affected,
with subsequent effects on the characteristics of the ensuing supernova
progenitors and explosions (Yoon et al. 2010; Eldridge et al. 2013; Podsiadlowski et al. 2004; Podsiadlowski 2014). In some systems, interaction can lead to the merger of the binary components, which could be the origin of rapid rotating massive stars (de Mink er al. 2013).

The effects of the optically thick winds are currently not accounted for in evolutionary models such as those presented here. As for the results from the evolutionary track, there is a great discrepancy in the temperature $\rm T_{\ast}$ between observations and evolutionary models.
However, this effect is well known: the radius $\rm R_{2/3}$ where the optical depth $\rm \tau=\frac{2}{3}$ is reached is much larger than the radius of the star without accounting for the optical thickness of the wind.  Thus, the observed $\rm T_{eff}$ is lower than those of the model calculations (Schaller et al. 1992;  Meyent \& Maeder 2003; Hamann et al. 2006; Crowther 2007). The observed late-type WC stars with low luminosity are found to have much larger radii and lower surface temperatures than those predicted by theoretical models .

The distance has an impact on the absolute quantities derived from spectral analysis, such as luminosity and mass-loss rate.
However, most distances to the WR stars in the Galactic system were not well constrained before Gaia. This is because these stars are too far to have a reliable HIPPARCOS parallax; the unobscured sources in particular are often relatively isolated. Recently, trigonometric parallax became available for the first time from the Gaia satellite (Hamann et al. 2019; Sander et al. 2012). On average, the new distances are smaller by only $10\%$ than the values adopted in previous work (Hamann et al. 2006). Based on Gaia DR2, Sander et al. (2019) re-examined a previously studied WC star sample to derive key properties of the Galactic WC population.
The newly obtained luminosities are found in the range of $\log L/L_{\odot}=4.9-6.0$ with one outlier (WR 119) having $\log L/L_{\odot}=4.7$. For example, WR 59 underwent the most drastic luminosity revision with $\rm \log L/L_{\odot} = 5.76$, a shift by $\rm +0.86$ dex. These new results provide convenient conditions with which to study the evolution of WC stars.

In this paper, we intend to explore whether the close binary evolution scenario succeeds in producing low-luminosity WC stars.

Specifically, we explore the consequences of close binary scenario on the evolution of the strong mass loss induced by mass transfer in RLOF events occurring at various evolutionary stages, and analyze the potential impact of such close binary evolution on the populations of supergiants and WR stars. Finally, we discuss the evolutionary properties of the progenitor of low-luminosity WC stars originating from such evolutionary histories.

In Section 2, we describe the various parameters in the models.
In Section 3, the results of numerical calculations on the evolution of single stars and binary systems are analyzed in detail.
In Section 4, we provide the necessary discussion of the validity of the formation of these WC stars in the binary system. Finally, we provide conclusions and a summary in Section 5. The effect of accretion efficiency on the evolution of the binary system is investigated in the Appendix.

\section{The initial parameters and model descriptions}
\begin{table}
\scriptsize{
\caption{The parameters adopted in our calculations. The meaning of each column is as follows$^{*}$. }
\begin{center}
\begin{threeparttable}
\setlength{\tabcolsep}{1mm}
\begin{tabular}{lrrrrrrrrrr}
\hline\hline
Models & $M_{\rm 1,ini}$& $M_{\rm 2,ini}$ &
$V_{\rm 1,ini}$ & $V_{\rm 2,ini}$  & $P_{\rm orb,ini}$& $\rm \beta$ & The wind formalism & \\\hline
  & $M_{\odot}$& $M_{\odot}$ &
km/s & km/s & days &
 & &\\

\hline
S1  &30   &..   &0   &..     &..           &..&Dutch &\\
S2  &30   &..   &300 &..     &..           &..& Dutch &\\
S3  &20   &..   &300 &..     &..           &..& Dutch &\\
S4  &40   &..   &300 &..     &..           &..& Dutch &\\
S5  &35   &..   &300 &..     &..           &..& Dutch&\\
S6  &30   &..   &300 &..     &..           &..&Van and Pot& \\

B1  &30  &27 &300  &300   &6.00       & 0.4&Dutch&\\
B2  &30  &27 &300 &300  &20.0       &0.4 &Dutch&\\
B3  &30  &27 &300 &300  &1000.0     & 0.4  &Dutch&\\
B4  &20  &16 &300 &300  &1000.0     & 0.4 &Dutch&\\
B5  &30  &27 &300 &300  &500.0      &0.4 &Dutch&\\
B6  &30  &27 &300 &300  &1000.0     &0.8&Dutch&\\
B7  &30  &27 &300 &300  &1000.0     &0.2 &Dutch&\\
B8  &30  &27 &300 &300  &1000.0     &0.0 &Dutch&\\
B9  &30  &27 &300   &300    &6.00       & 0.0&Dutch&\\
B10  &30  &27 &300  &300    &6.00       & 0.8&Dutch&\\

\hline\hline
\end{tabular}
 \begin{tablenotes}
        \footnotesize
        \item[*]
The symbol S denotes single stars whereas the symbol B denotes the evolution of the binary system.
$M_{\rm 1,\rm ini}$: the initial mass of the primary star in unit of $M_{\odot}$; $M_{\rm 2,\rm ini}$:
the initial mass of the secondary star in units of $M_{\odot}$; $\rm V_{\rm 1,\rm ini}$: the initial equatorial
velocity of the primary star in units of $Km/s$; $V_{\rm 2, \rm ini}$: the initial equatorial velocity of the
secondary star in unit of $\rm km/s$; $P_{\rm orb,\rm ini}$: the initial orbital period; Dutch: the wind mass-loss formalism by Dutch scheme; Van and Pot: the wind mass-loss formalism is same as Dutch scheme except for
the RSG wind formalism proposed by Vanbeveren \& Mennekens (2015) and the WR wind formalism which derived by Potsdam group.
      \end{tablenotes}
    \end{threeparttable}
\end{center}
}
\end{table}

\begin{table*}[h]
\begin{center}
\scriptsize{
\caption{Major evolutionary parameters for five models including single stars and the primary star in binaries. }
\begin{threeparttable}
\begin{tabular}{llrclllllllllllllcll}
  & & & & & & & &  & &\\
\hline\hline Sequence&Age, & $M_1$ & $\rm \log(\frac{R}{R_{\odot}})$ &$\log T_{\rm eff}$& $\log (\frac{L}{L_{\odot}})$&$\log T_{\rm c}$& $\log \rho_{\rm c}$& $\rm \frac{N}{N_{\rm ini}}$ & $V_{\rm eq}$\\\hline
  &Myr&$M_{\odot}$& & K&  &K&$\rm g/cm^{3}$&&km/s \\
\hline
ZAMS&&&&&&&&&\\
S1&0.000&30.000&0.870&4.596&5.077&7.562&0.496&1.00&0.00\\
S2&0.000&30.000&0.882&4.583&5.048&7.557&0.485&1.00&300.00\\
B1&0.000&30.000&0.872&4.590&5.057&7.571&0.528&1.00&300.00\\
B2&0.000&30.000&0.872&4.590&5.057&7.571&0.528&1.00&300.00\\
B3&0.000&30.000&0.872&4.590&5.057&7.571&0.528&1.00&300.00\\
\hline
ECHB&&&&&&&&&\\
S1&6.368&26.917&1.640&4.305&5.454&7.780&1.074&1.00&0.00\\
S2&6.724&25.998&1.638&4.310&5.467&7.781&1.071&5.22&4.52\\
B1&6.638&17.967&1.321&4.450&5.395&7.778&1.104&11.75&107.40\\
B2&6.840&25.759&1.644&4.309&5.476&7.783&1.072&6.04&74.17\\
B3&6.955&25.809&1.562&4.352&5.486&7.782&1.065&6.95&4.78\\
\hline
ECHEB&&&&&&&&&\\
S1&6.877&14.105&0.145&5.115&5.703&8.943&4.976&12.58&0.00\\
S2&7.244&13.705&-0.166&5.264&5.676&8.944&4.990&18.49&248.04\\
B1&7.179&8.944&-0.418&5.329&5.434&8.937&5.082&0.00&1.83\\
B2&7.349&10.340&-0.409&5.350&5.537&8.939&5.017&0.00&0.73\\
B3&7.473&12.975&-0.309&5.332&5.662&8.933&4.922&11.89&180.77\\
\hline
ECCB&&&&&&&&&\\
S1&6.877&14.104&0.131&5.122&5.705&9.077&5.731&12.58&0.00\\
S2&7.244&13.705&-0.184&5.274&5.681&9.080&5.779&18.48&235.27\\
B1&7.179&8.944&-0.419&5.330&5.435&9.002&5.618&0.00&1.79\\
B2&7.349&10.340&-0.423&5.359&5.542&9.014&5.558&0.00&0.73\\
B3&7.473&12.974&-0.361&5.361&5.675&9.031&5.451&11.89&187.83\\
\hline
BROLF1&&&&&&&&&\\
B1&5.971&27.933&1.349&4.435&5.391&7.641&0.643&3.64&160.23\\
B2&6.848&25.702&1.721&4.275&5.494&8.100&2.108&6.041&47.85\\
B3&6.967&25.724&2.826&3.725&5.504&8.232&2.537&7.00&0.70\\
\hline
EROLF1&&&&&&&&&\\
B1&6.386&18.427&1.345&4.426&5.345&7.664&0.766&10.21&115.43\\
B2&6.856&16.588&1.780&4.278&5.626&8.253&2.590&12.53&22.76\\
B3&6.977&19.869&2.863&3.735&5.619&8.275&2.659&8.64&0.70\\
\hline
BROLF2&&&&&&&&&\\
B1&6.453&18.393&1.366&4.418&5.356&7.672&0.789&11.43&121.25\\
\hline
EROLF2&&&&&&&&&\\
B1&6.614&17.983&1.368&4.424&5.383&7.725&0.944&11.73&118.71\\
\hline
BROLF3&&&&&&&&&\\
B1&6.646&17.924&1.392&4.429&5.452&8.069&2.044&11.75&29.82\\
\hline
EROLF3&&&&&&&&&\\
B1&6.663&14.283&1.420&4.459&5.631&8.275&2.691&12.60&66.40\\
  & & & & & &  & & & &\\
\hline\hline
\end{tabular}
 \begin{tablenotes}
        \footnotesize
        \item[]Abbreviations: ZAMS-zero age main sequence; TAMS-the terminal of core hydrogen burning; ECHEB-the end of core helium burning; ECCB-the end of core carbon burning. BROLF1-the beginning of the first episode of RLOF, EROLF1-the end of the first episode of RLOF, BROLF2-the beginning of the second episode of RLOF, EROLF2-the end of the second episode of RLOF, BROLF3-the beginning of the third episode of RLOF, EROLF3-the end of the third episode of RLOF.
      \end{tablenotes}
    \end{threeparttable}
\label{tablemod}
}
\end{center}
\end{table*}

\begin{table*}[h]
\begin{center}
\scriptsize{
\caption{Surface chemical abundances for $\rm ^{1}H$, $\rm ^{4}He$, $\rm ^{12}C$, $\rm ^{14}N$, $\rm ^{16}O$ and $\rm ^{22}Ne$ in mass fraction, and three abundance ratios (in number) $\rm \frac{^{22}Ne}{^{20}Ne}$, $\rm \frac{^{14}N}{^{12}C}$, and $\rm \frac{^{14}N}{^{16}O}$ for single stars and the primary star in the binary system at selected evolutionary points. Other abbreviations are the same as table 2.}
\begin{tabular}{llllllllllllccccccccccccccc}
  & & & & & & & &  & &\\
 \hline\hline Sequence & $t$(Myr), & $ X_{\rm ^{1}H}$, & $Y_{\rm ^{4}He}$, & $\rm \log (^{12}C)$,  & $\rm \log (^{14}N)$, & $\rm \log (^{16}O)$, & $ \rm \log (^{22}Ne)$, & $ \rm \log ( \frac{^{22}Ne}{^{20}Ne})$, &
$\rm \log ( \frac{^{14}N}{^{12}C})$, & $\rm \log (\frac{^{14}N}{^{16}O})$ \\
\hline
 & & & & & &  & &  & &\\
ZAMS&&&&&&&&&&&\\
S1&0.000&0.725&0.261&-2.623&-3.153&-2.185&-4.039&-1.134&-0.597&-0.91\\
S2&0.000&0.725&0.261&-2.623&-3.153&-2.185&-4.039&-1.134&-0.597&-0.91\\
B1&0.000&0.725&0.261&-2.623&-3.153&-2.185&-4.039&-1.134&-0.597&-0.91\\
B2&0.000&0.725&0.261&-2.623&-3.153&-2.185&-4.039&-1.134&-0.597&-0.91\\
B3&0.000&0.725&0.261&-2.623&-3.153&-2.185&-4.039&-1.134&-0.597&-0.91\\
\hline
ECHB&&&&&&&&&&&\\
S1&6.368&0.725&0.261&-2.623&-3.153&-2.185&-3.960&-1.134&0.253&-0.91\\
S2&6.724&0.702&0.284&-2.951&-2.436&-2.319&-4.276&-1.449&2.806&-0.06\\
B1&6.638&0.575&0.411&-4.022&-2.083&-3.016&-6.209&-3.301&1.872&0.99\\
B2&6.840&0.692&0.294&-3.003&-2.372&-2.365&-4.421&-1.515&0.564&0.05\\
B3&6.955&0.675&0.311&-3.062&-2.311&-2.424&-4.494&-1.588&0.684&0.17\\
\hline
ECHEB&&&&&&&&&&&\\
S1&6.877&0.183&0.804&-4.002&-2.053&-3.539&-7.080&-4.241&76.114&1.54\\
S2&7.244&0.000&0.972&-2.222&-1.886&-2.524&-3.396&-0.551&1.859&0.69\\
B1&7.179&0.000&0.347&-0.297&-17.917&-0.886&-1.868&1.042&-17.688&-16.97\\
B2&7.349&0.000&0.263&-0.285&-17.676&-0.700&-1.872&1.006&-17.459&-16.92\\
B3&7.473&0.000&0.960&-1.699&-2.078&-2.219&-3.119&-0.196&-0.446&0.20\\
\hline
ECCB&&&&&&&&&&&\\
S1&6.877&0.182&0.804&-4.002&-2.053&-3.539&-7.080&-4.242&76.111&1.54\\
S2&7.244&0.000&0.972&-2.220&-1.886&-2.523&-3.395&-0.550&1.849&0.69\\
B1&7.179&0.000&0.347&-0.297&-17.917&-0.886&-1.868&1.042&-17.687&-16.97\\
B2&7.349&0.000&0.263&-0.285&-17.676&-0.699&-1.872&1.006&-17.458&-16.92\\
B3&7.473&0.000&0.960&-1.699&-2.078&-2.219&-3.119&-0.196&-0.446&0.20\\
\hline
BROLF1&&&&&&&&&&&\\
B1&5.971&0.713&0.273&-2.809&-2.592&-2.258&-4.223&-1.318&0.150&-0.28\\
B2&6.848&0.692&0.294&-3.003&-2.372&-2.365&-4.421&-1.515&0.564&0.05\\
B3&6.967&0.675&0.311&-3.067&-2.308&-2.427&-4.499&-1.592&0.691&0.18\\
\hline
EROLF1&&&&&&&&&&&\\
B1&6.386&0.621&0.366&-3.616&-2.144&-2.705&-5.067&-2.160&1.405&0.62\\
B2&6.856&0.342&0.645&-3.961&-2.055&-3.499&-6.709&-3.796&1.839&1.50\\
B3&6.977&0.594&0.392&-3.205&-2.217&-2.563&-4.659&-1.751&0.921&0.41\\
\hline
BROLF2&&&&&&&&&&&\\
B1&6.453&0.593&0.394&-3.941&-2.095&-2.926&-5.774&-2.866&1.780&0.89\\
\hline
EROLF2&&&&&&&&&&&\\
B1&6.614&0.577&0.410&-4.020&-2.084&-3.011&-6.187&-3.279&1.869&0.99\\
\hline
BROLF3&&&&&&&&&&&\\
B1&6.646&0.575&0.412&-4.023&-2.083&-3.018&-6.216&-3.308&1.873&0.99\\
\hline
EROLF3&&&&&&&&&&&\\
B1&6.663&0.233&0.753&-3.975&-2.053&-3.585&-6.522&-3.607&1.855&1.59\\
\hline\hline
  & & & & & &  & & & &\\
\end{tabular}
\label{tablemod}
}
\end{center}
\end{table*}

The models presented in this investigation are calculated
with the MESA code (Paxton et al. 2011, 2013, 2015, 2018).
We use the Schwarzschild criterion to determine the boundaries
of the convective region. The mixing
length is $l_{\rm m}=1.5 H_{\rm P}$, where $H_{\rm P}$ is the pressure scale height at the outer boundary of the core. We consider convective core-overshooting adopting,
an overshooting parameter of 0.335 pressure scale heights following Brott et al. (2011). This value was
calibrated by adopting observational constraints of Large Magellan Cloud stars with masses of about $15 M_{\odot}$, making it a more appropriate choice in contrast to
smaller values derived from intermediate-mass eclipsing
binaries (Stancliffe et al. 2015).
Our initial models consist of one or two zero-age main sequence (ZAMS) stars of Population I ($X_{\rm H}=0.72, Z=0.014$).

The quantity $\beta$ is the faction of the mass transfer rejected by the accretor and $\rm \beta=0.0$ denotes the conservative case (where all the transferred mass via RLOF is obtained by the gainer).
The efficiency of mass transfer is chosen as $1-\beta=0.6$ in the referenced model.  The remaining transferred mass which carries the
specific orbital angular momentum of the mass gainer is directly expelled from the system. The matter that is transferred to the secondary carries a significant amount of angular momentum, especially if it passes through an accretion disk. This can bring the secondary to break-up rotation after accreting only about 10$\%$ of its original mass (Packet 1981). The secondary then has to get rid of some or most of its angular momentum before further accretion can take place. This angular momentum catastrophe therefore potentially limits the amount of accretion. This process necessarily leads to nonconservative mass transfer, and a small mass-transfer efficiency. Although Vanbeveren (1991) proposed that the accretion efficiency depends on the mass ratio of the system, a value of 0.5 has been commonly adopted so far, regardless of the physical mechanism behind the mass loss or the properties of the binary system (De Loore \& De Greve 1992; Vanbeveren  et al. 1998a, b c). Also, we adopt an interesting variant of Jeans's mode in this paper which is isotropic re-emission. The transferred matter in the vicinity of the accretor is ejected as a fast, isotropic wind (Soberman et al. 1997). Mass shedding of accreted material from the equator of the accreting star leads to the assumption that the remainder of the mass is lost from the system taking away the specific orbital angular momentum of the accreting star. This process is similar to what is believed to be the cause of the Be phenomenon when the rotation rate is close to the critical limit (e.g., Owocki 2006).
The effect of accretion efficiency in two cases of mass transfer (Case A and Case C) is investigated in the Appendix.

We use the Dutch scheme in MESA for
both hot and cool wind mass-loss rates, with the Dutch scaling
factor of 1.0 \footnote{The Dutch wind mass-loss scheme is a combination of the prescriptions
of Vink et al. (2001) (when $\rm T_{eff} \geq 10^{4} $ K and $\rm X_{surf} \geq
0.4$ ), Nugis \& Lamers (2000) (when $\rm T_{eff} \geq 10^{4} $ K and $\rm X_{surf} <
0.4$ ), and de Jager et al. (1988) (when $\rm T_{eff} < 10^{4} $ K)}. Radiative opacities were interpolated from the OPAL tables (Iglesias \&
Rogers 1996). The opacity increase due to Fe-group elements at $\rm T \sim 180 $ kK plays an
important role in determining the envelope structure in our stellar models.
The Potsdam group reconsidered the mass-loss rates of Galactic WN, WC, and WO stars and the impact of revised distances from Gaia DR2 (Hamann et al. 2019; Sander et al. 2019). In order to investigate the effect of stellar winds on the evolution of WC stars, we also included both the RSG wind formalism proposed by Vanbeveren \& Mennekens (2015) and the WR mass-loss rate formalism derived by Potsdam group in the single-star model S6.

We take into account various instabilities induced
by rotation that result in mixing: Eddington-Sweet circulation,
dynamical and secular shear instability, and the Goldreich-Schubert-Fricke
instability.
The rotational mixing due to these hydrodynamical
instabilities is treated as a diffusive process,
following Heger et al. (2000). The diffusion coefficients are used for the transport of both
the angular momentum and
the chemical species.
The contribution of the rotationally
induced instabilities to the total diffusion coefficient is
reduced by a factor of $\rm f_{c}=0.0228$. This factor has been
calibrated to reproduce the observed nitrogen surface abundances as a function of the projected rotational velocities for
stars in the Large Magellanic Cloud sample (NGC 2004) of the
FLAMES survey (Brott et al. 2011). The inhibiting effect
of chemical gradients on the efficiency of rotational mixing processes
is regulated by the parameter $\rm f_{\mu}$.
We adopt a value $\rm f_{\mu}=0.1$ as in Yoon et al. (2006) who calibrated this
parameter to match the observed surface helium abundances.

The initial parameters for single stars and the binary system are listed in Table 1. The binary orbit is assumed to be circular and the Roche lobe radius is given by the formula proposed in Eggleton (1983).
We chose three initial orbital periods corresponding to cases where the first mass-transfer event occurs during the MS phase (6.0 days, Case A) after core H exhaustion but before the He ignition in the core (20 days, Case B), during the core He-burning (1000 days, Case C).

\section{Results of numerical calculations}
We present nonrotating and rotating single-star models to compare them with binary star models. We focus our investigation on the evolution of the primary star and explore whether binary evolution via Case A, Case B, or Case C mass transfer could give rise to diverse WR stars in terms of the amount of removed hydrogen or helium envelope. In all models, we calculate the evolution until at least the end of central carbon burning. The evolutionary track of the primary in the close binary system composed of a 30 $M_{\odot}$ star and a 27 $M_{\odot}$ star is shown for cases where the initial orbital period is equal to 6.0, 20 and 100 days.

We adopt the definitions of various kinds of stars according to the relationship between spectral type and surface abundance from Smith \& Maeder (1991) and Meynet \& Maeder (2003), as follows:

-O stars have a surface hydrogen mass fraction $\rm X_{H} > 0.3$ and $\rm \log (T_{\rm eff}/K) > 4.5$;

-B stars have typically $\rm 4.0 <\log (T_{\rm eff}/K)<4.5 $;

-Red supergiant stars (RSGs) have $\rm \log (T_{\rm eff}/K) < 3.66$;

-Blue supergiant stars (BSGs) have $\rm \log (T_{\rm eff}/K) \geq 3.9$ and include O, B, and A stars;

-WR stars have $\rm \log( T_{eff}/K) > 4.0$  and $\rm X_{H} < 0.3$;

-WNL stars are WR stars with $\rm X_{H} > 10^{-5}$;

-WNE stars have $\rm X_{H }< 10^{-5}$ and a surface carbon abundance $\rm X_{C}$ lower than 0.1 times the surface nitrogen abundance $\rm X_{N}$;

-WC stars have $\rm X_{H } < 10^{-5}$ and $\rm \frac{X_{C}}{X_{N}}> 1.0$, and surface abundances (by number) such as $\rm \frac{C+O}{He}< 1.0$;

-WO stars have $\rm X_{H } < 10^{-5}$ and $\rm \frac{X_{C}}{X_{N}}> 1.0$, and  $\rm \frac{C+O}{He}>1.0$.

 The properties of single stars and the primary star in binaries, such as evolutionary age, actual mass, radius, effective temperature of the surface of the star, luminosity, central temperature and central density, ratio of the surface nitrogen to the initial value, and equatorial velocity, are presented in Table 2.

Surface chemical abundances for single stars and the primary star in binaries, such as evolutionary age, the logarithm of mass fraction of chemical elements such as $\rm ^{1}H$, $\rm ^{4}He$, $\rm \log (^{12}C)$, $\rm \log (^{14}N)$, $\rm \log (^{16}O)$, $\rm \log (^{22}Ne)$, and  the logarithm of three abundance ratios (in number) $\rm \log (\frac{^{22}Ne}{^{20}Ne})$, $\rm \log (\frac{^{14}N}{^{12}C})$, and $\rm \log (\frac{^{14}N}{^{16}O})$ are indicated in Table 3.

\begin{figure}
	\centering
  \includegraphics[width=0.5\textwidth]{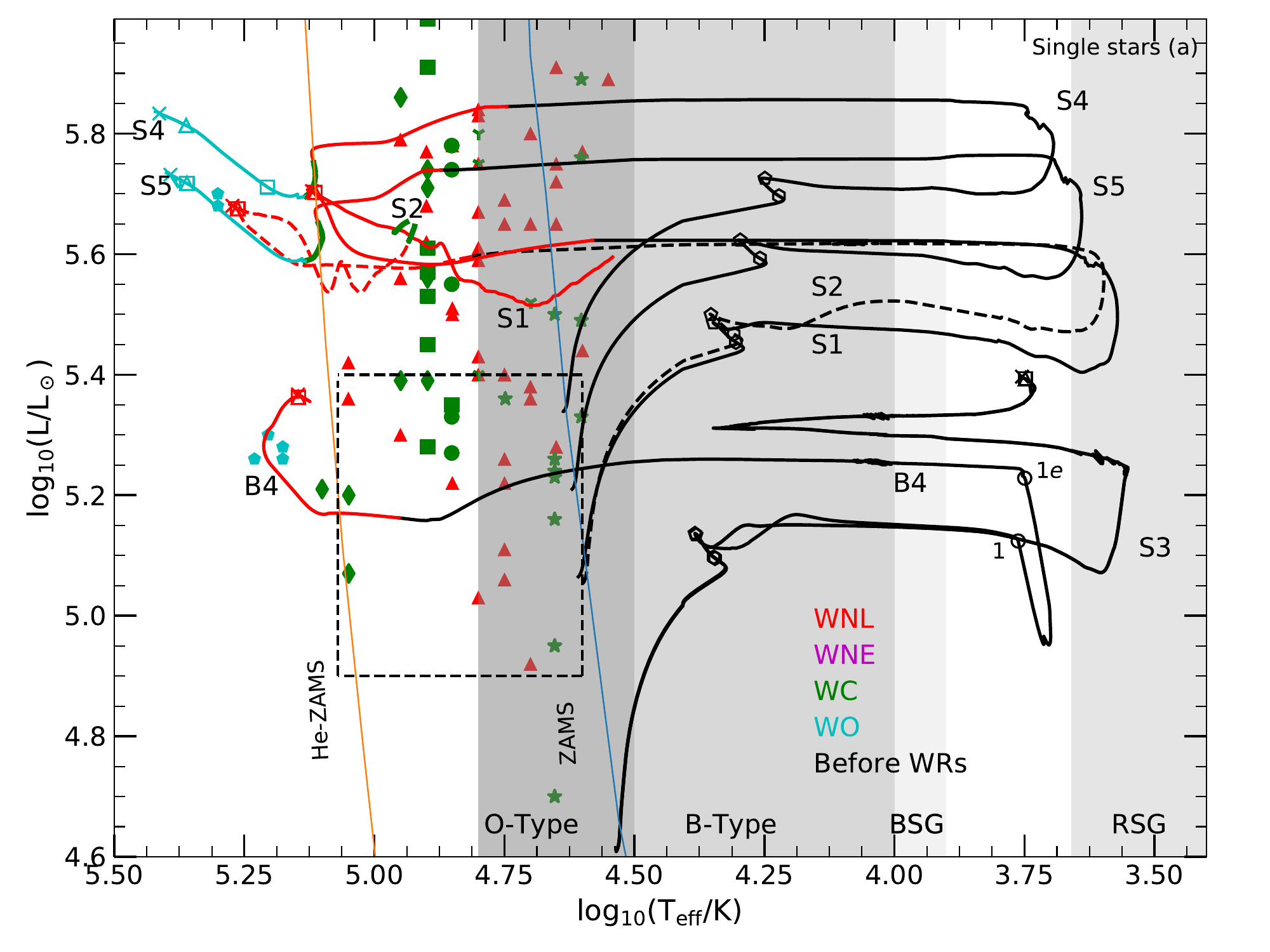}\\
  \includegraphics[width=0.5\textwidth]{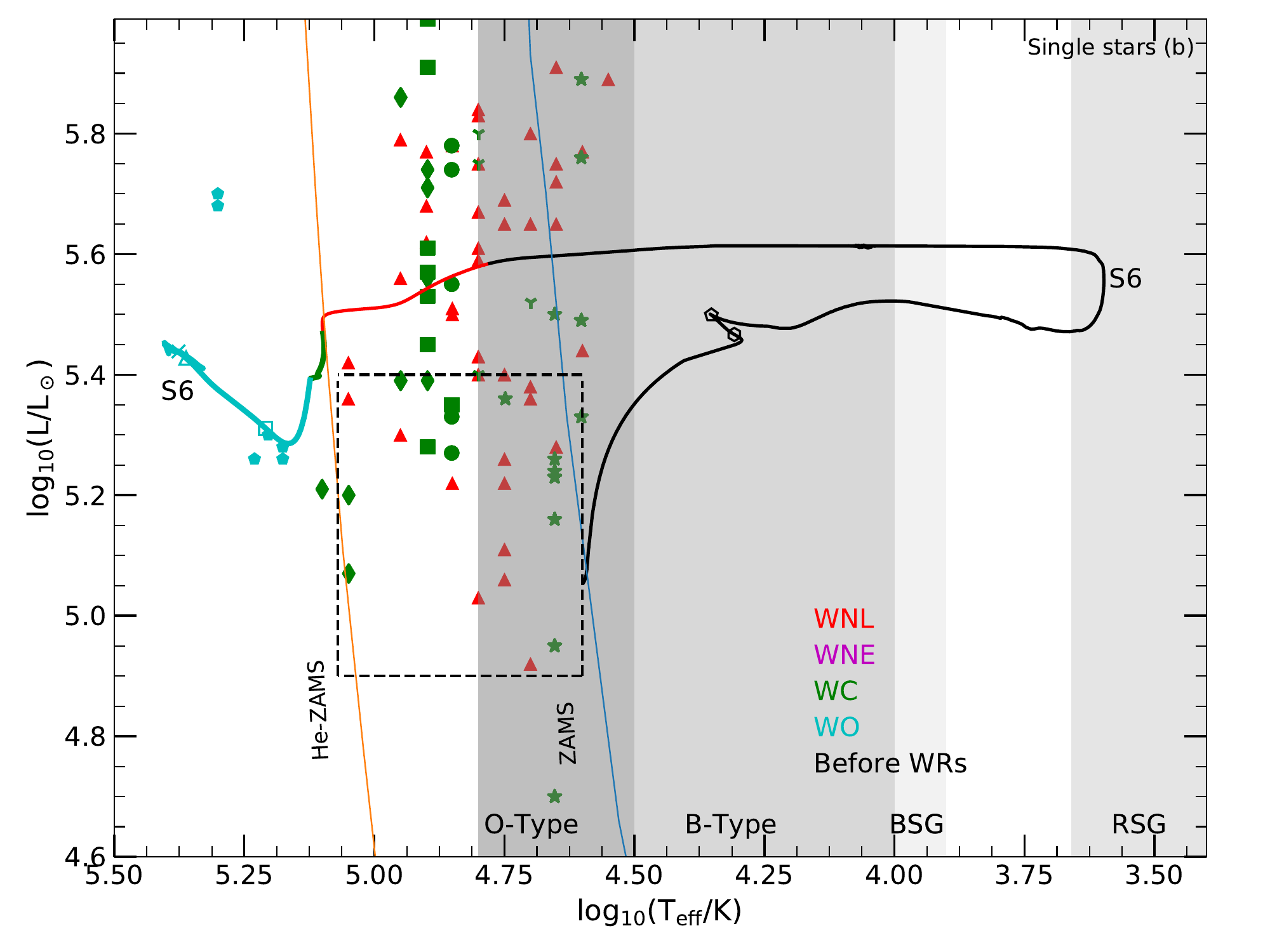}\\\
  \includegraphics[width=0.5\textwidth]{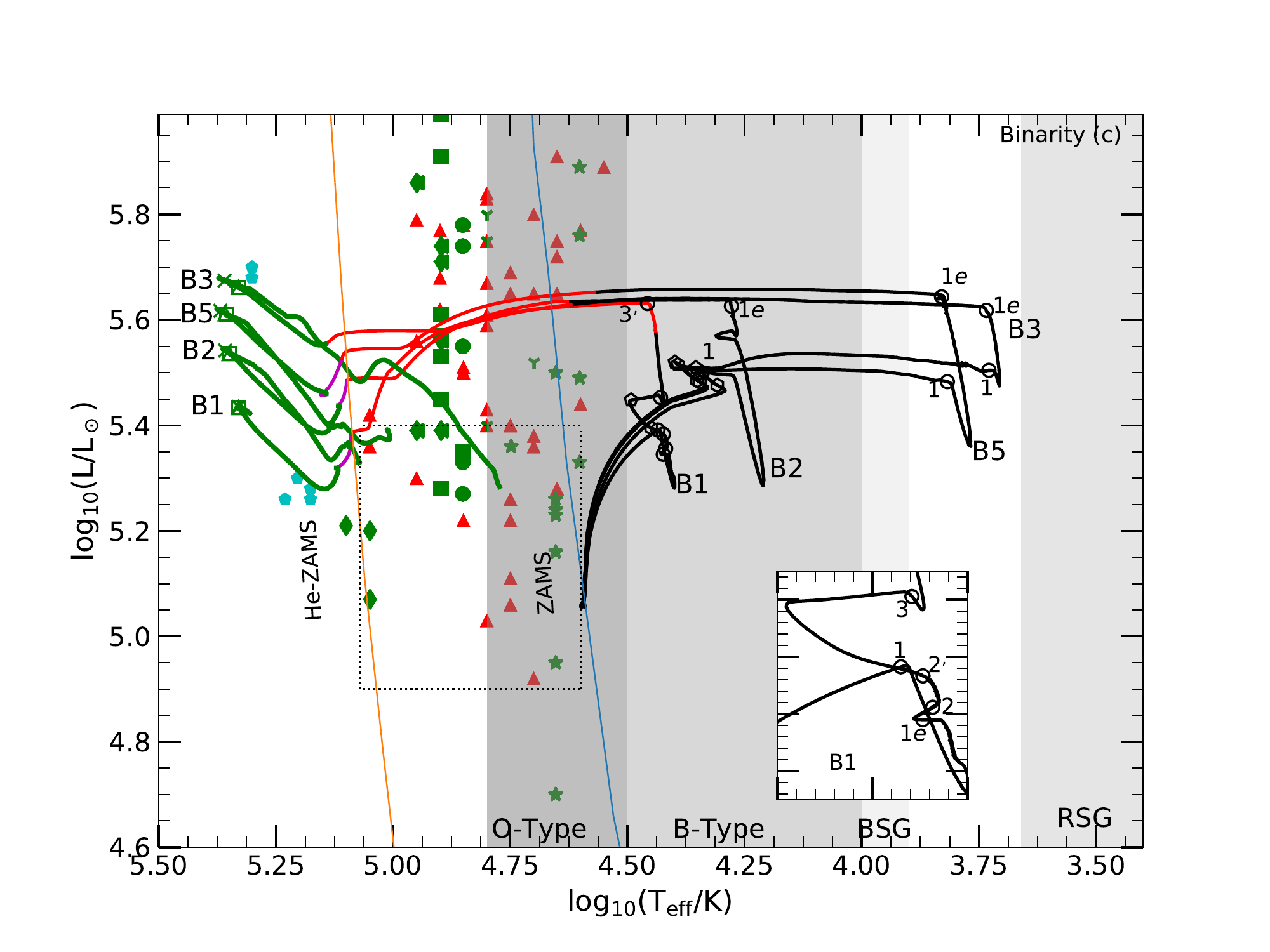}\\
  \caption{Hertzsprung-Russell diagram of the observed Galactic WC and WO stars with updated luminosities due to Gaia distances. The samples are taken from Sander et al. (2019).
Hertzsprung-Russell diagram with the evolutionary track of the massive models with the different types/phases marked in colours (Before WR stars: black; WNL:
red; WNE: pink; WC: green; WO: cyan). The range for the observed late WC stars with low luminosity is marked by a box.  Panel (a): single star models. Panel (b): single star model S6; Panel (c): rotating binary models.}
\label{Fig:parallax}
\end{figure}

\begin{figure}[h]
  \centering
  \includegraphics[width=0.5\textwidth]{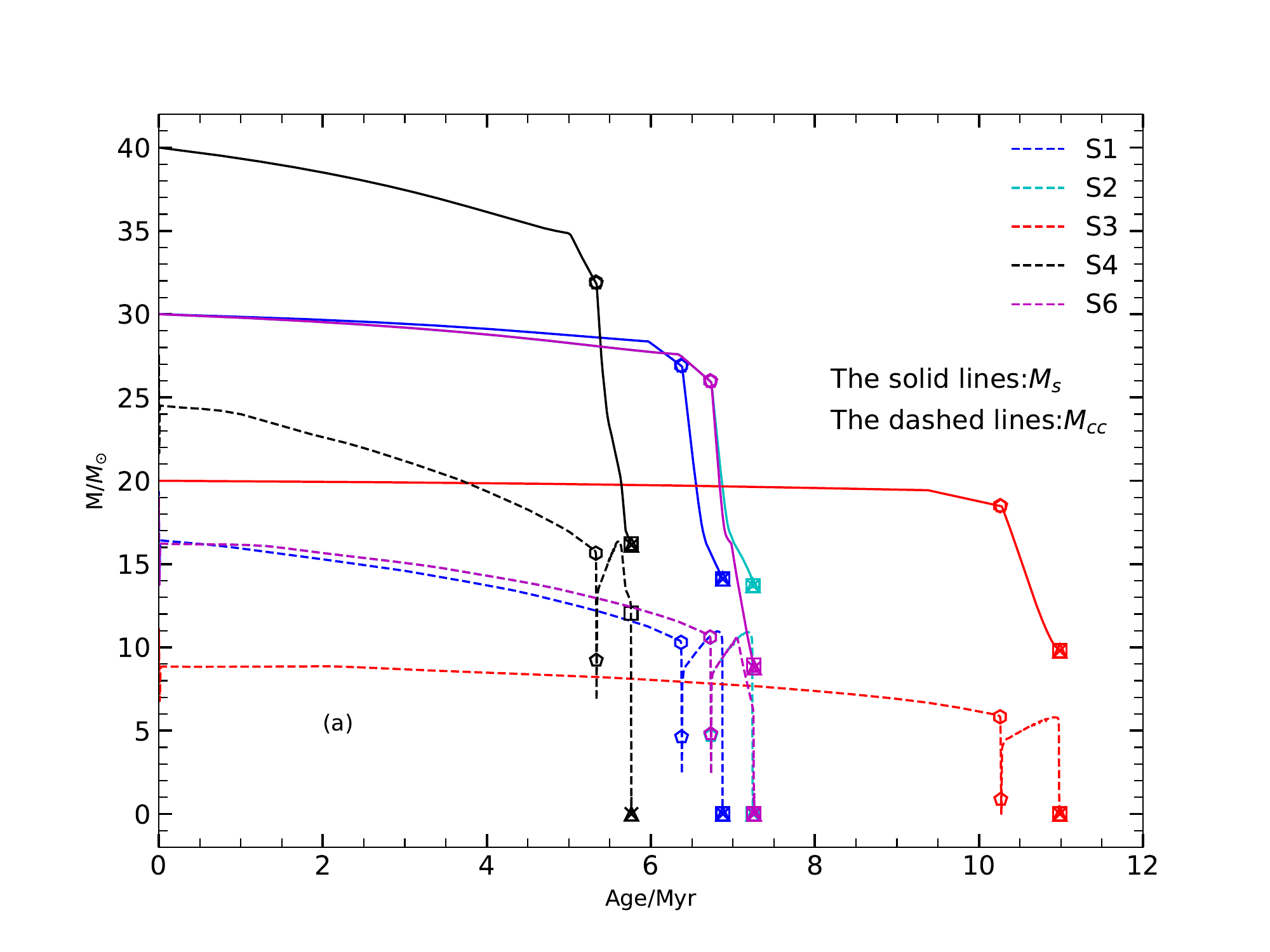}\\
  \includegraphics[width=0.5\textwidth]{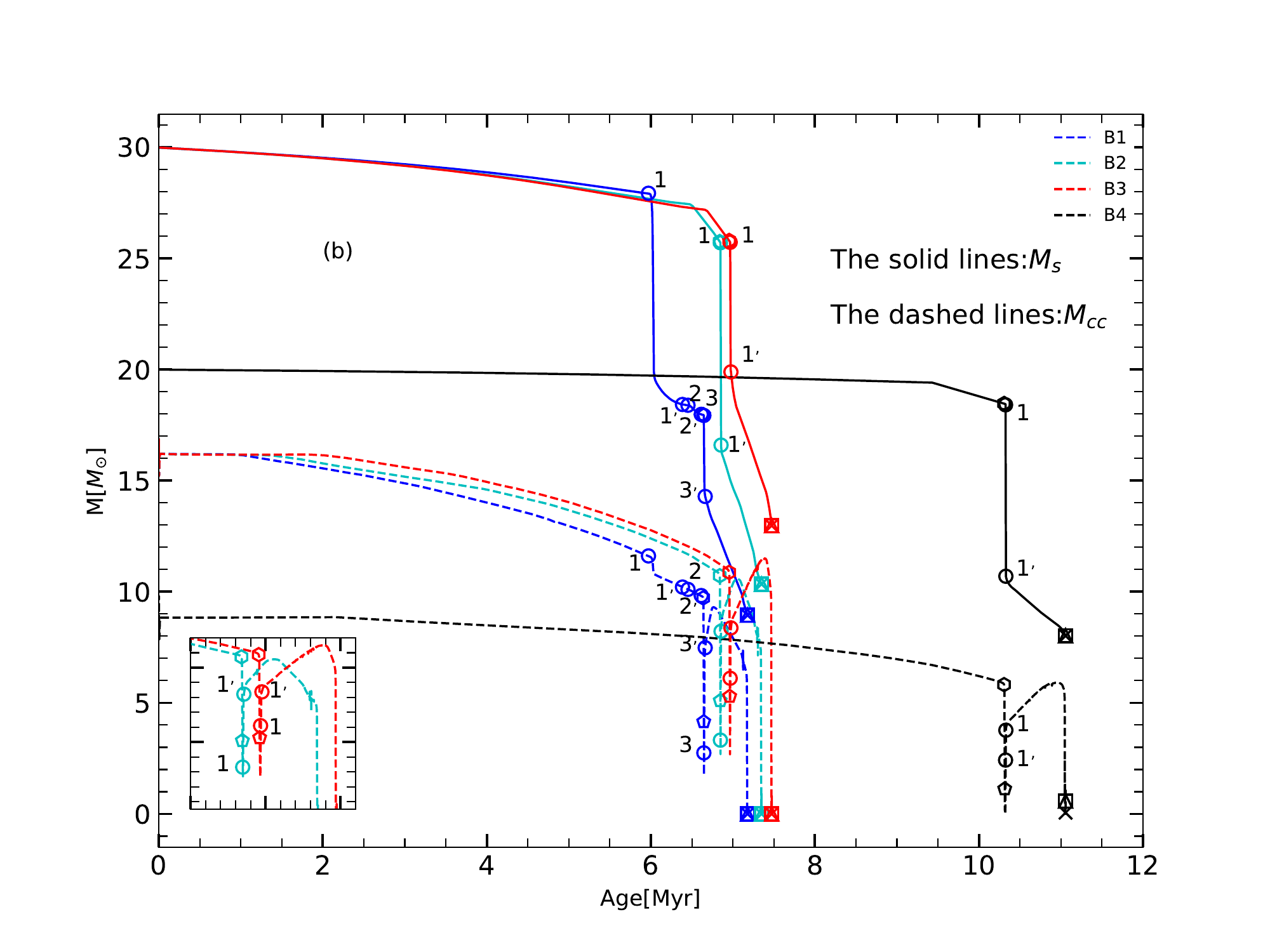}\\
  \caption{Convective core and stellar mass for nonrotating and rotating single stars and the primary star in rotating binaries vary as a function of evolutionary age.}
\end{figure}

\begin{figure}[h]
  \centering
  \includegraphics[width=0.5\textwidth]{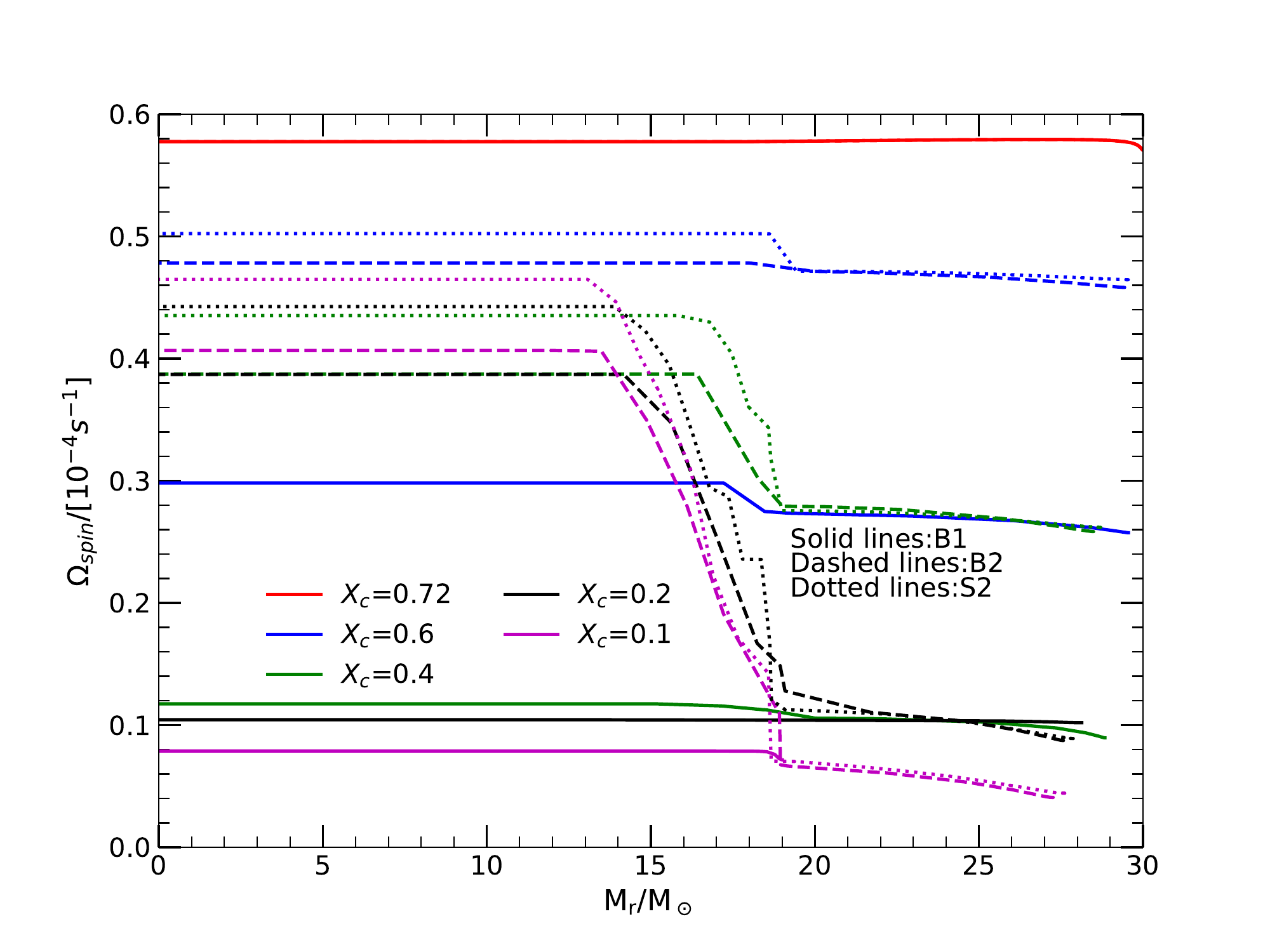}\\
  \caption{Rotation profiles as a function of the Lagrangian mass at various stages of MS evolution (labeled by the central hydrogen content $X_{c}$) of three models. The dotted lines indicate a rotating 30 $M_{\odot}$ single star S2. Dashed and solid lines indicate the primary star in B2 with initial $\rm P_{orb}=20.0$ days and the primary star in B1 with initial $\rm P_{orb}=6.0 $, respectively.}
\end{figure}

\begin{figure}[h]
  \centering
  \includegraphics[width=0.5\textwidth]{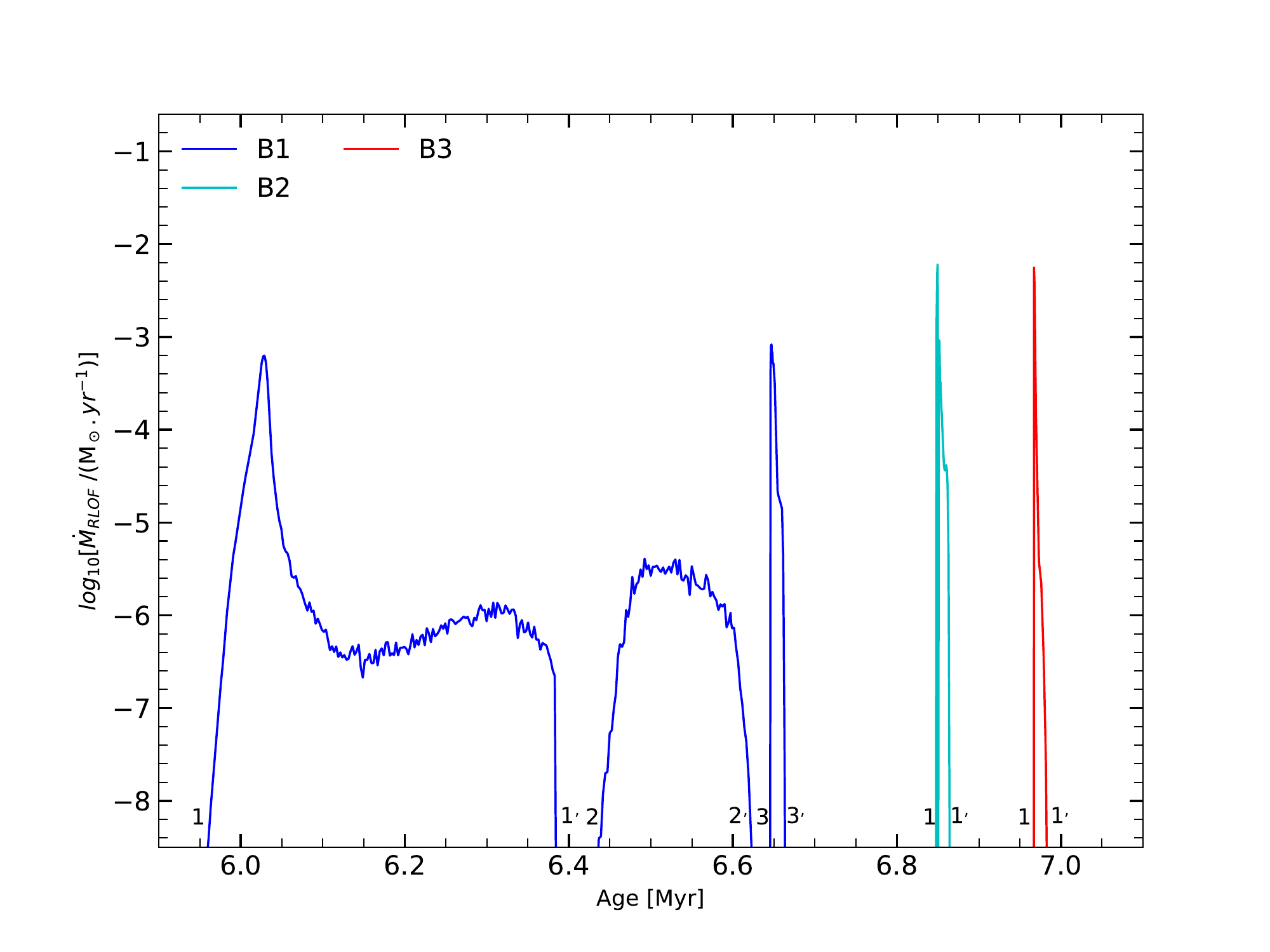}\\
  \caption{Mass-transfer rate due to RLOF as a function of time for three binaries B1, B2, and B3. }
\end{figure}

\begin{figure}[h]
  \centering
  \includegraphics[width=0.5\textwidth]{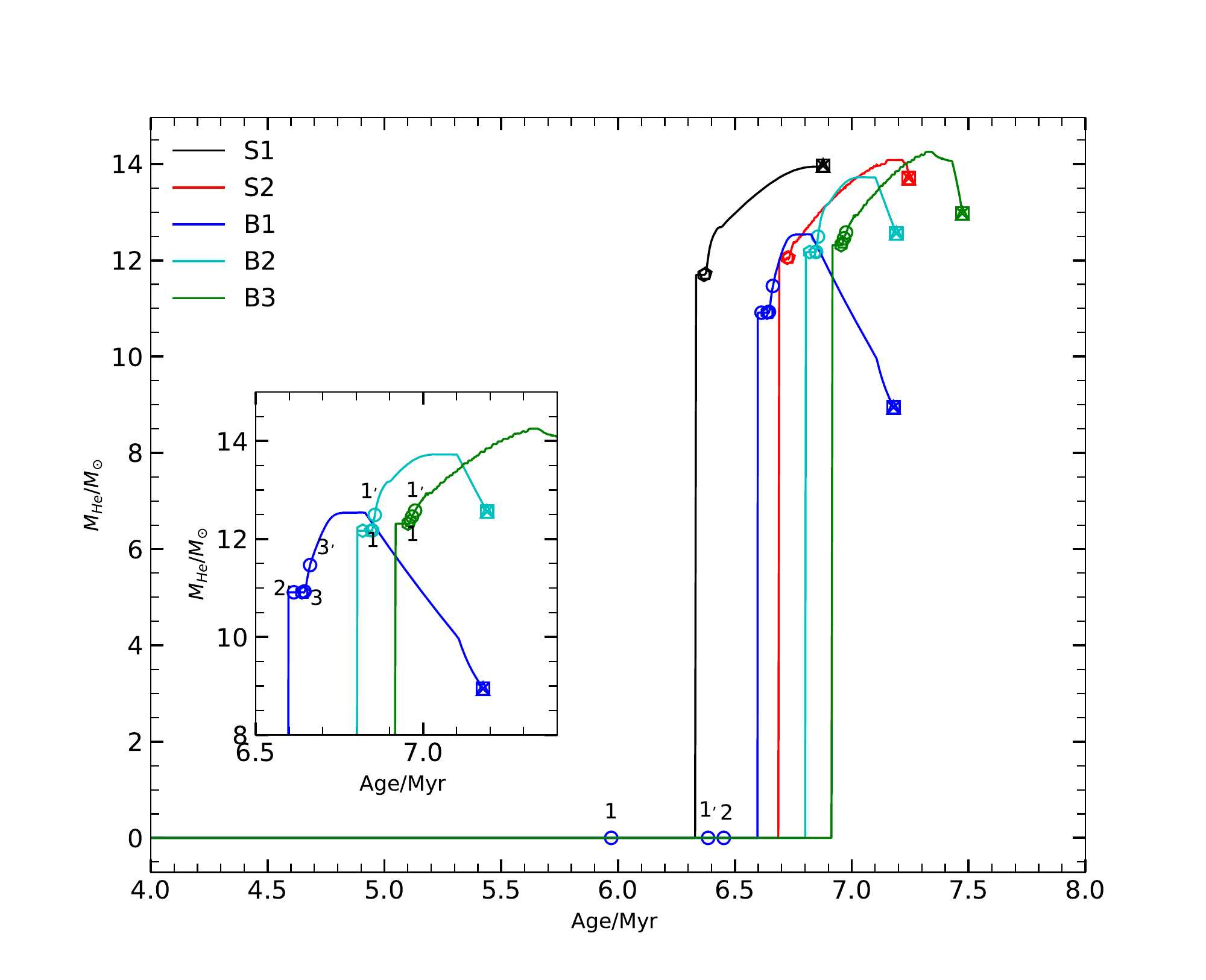}\\
  \caption{Helium cores for nonrotating and rotating single stars and the primary star in rotating binaries vary as a function of evolutionary age.}
\end{figure}

\begin{figure*}[h]
\centering
\includegraphics[width=0.35\textwidth]{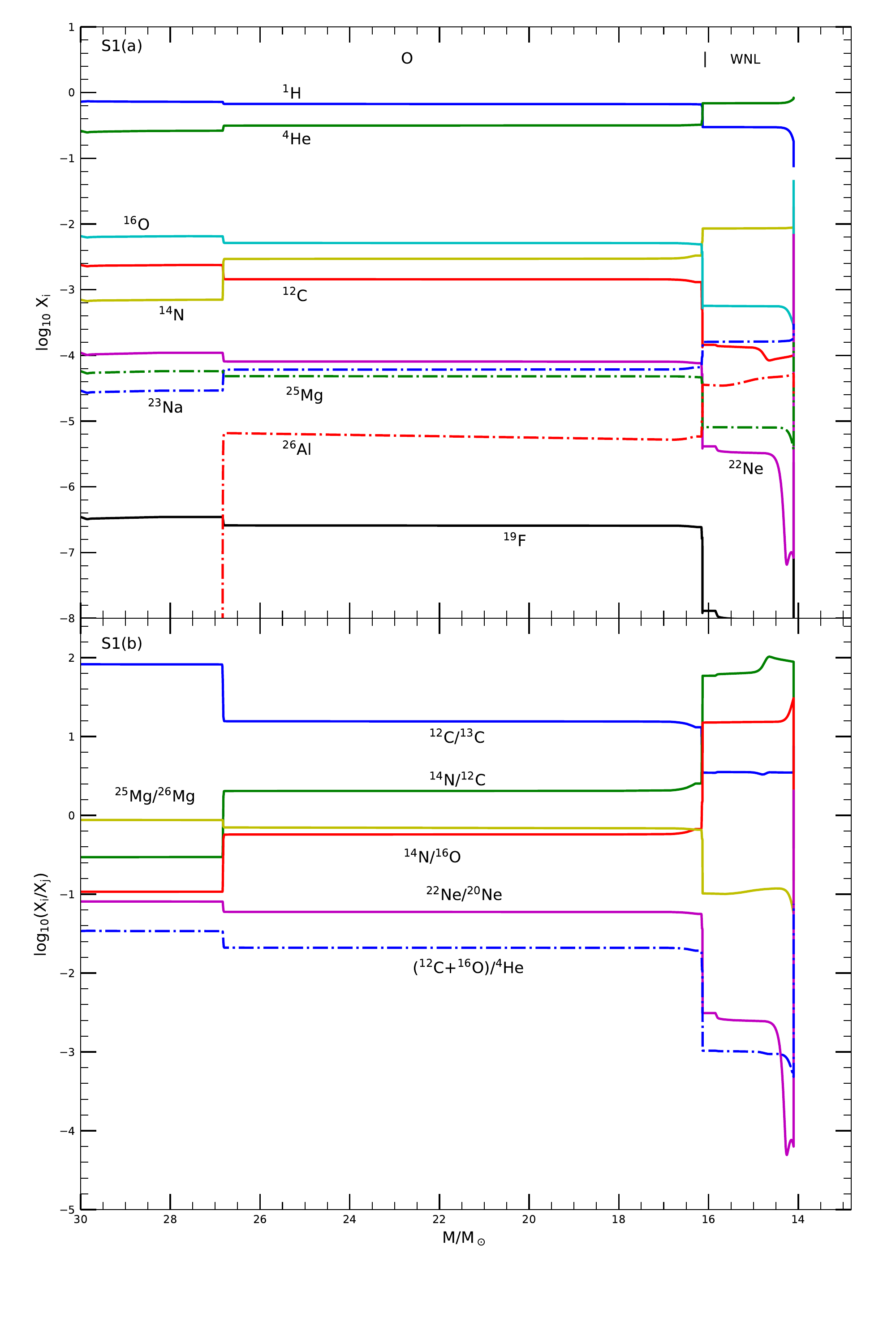}
\includegraphics[width=0.35\textwidth]{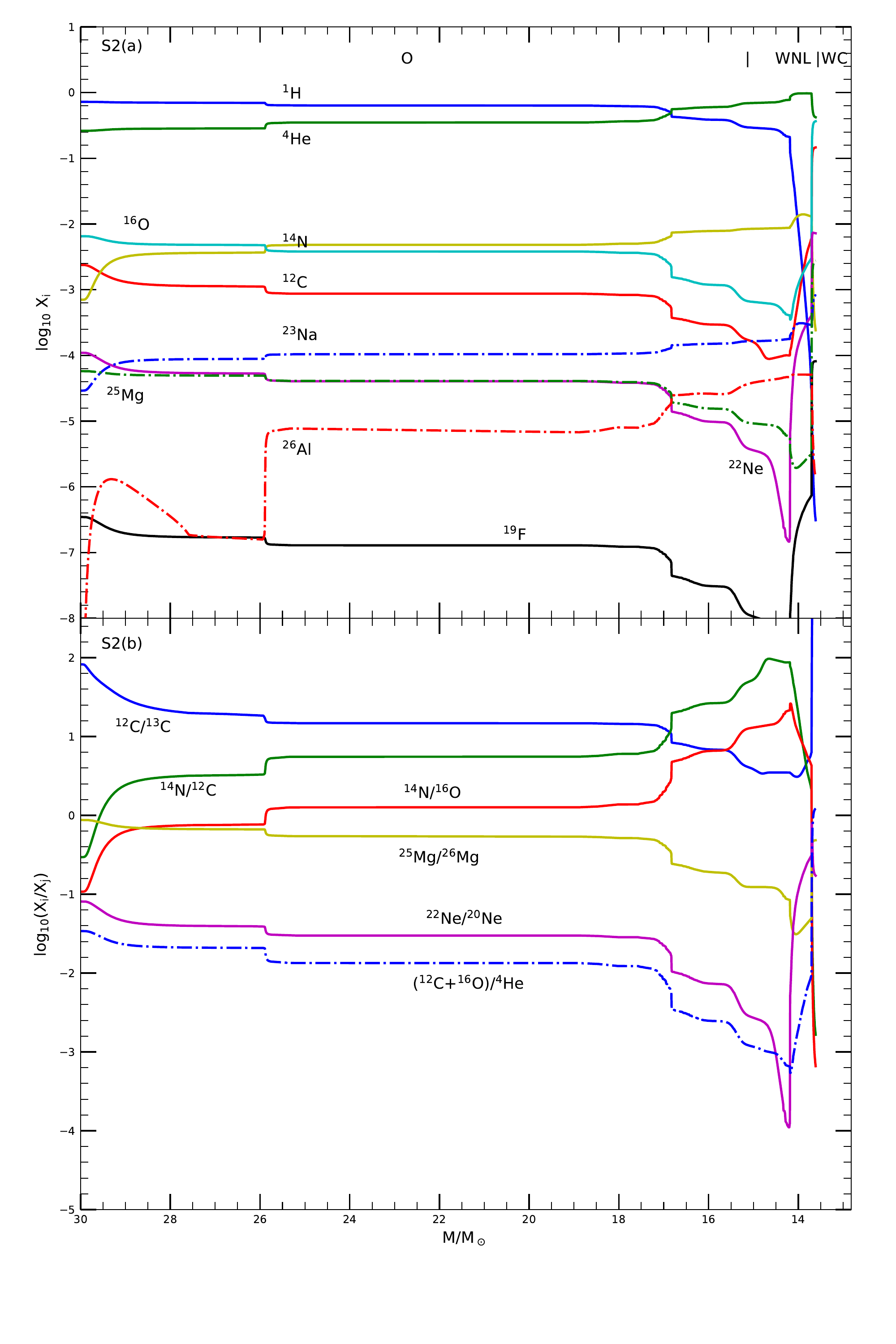}
\includegraphics[width=0.35\textwidth]{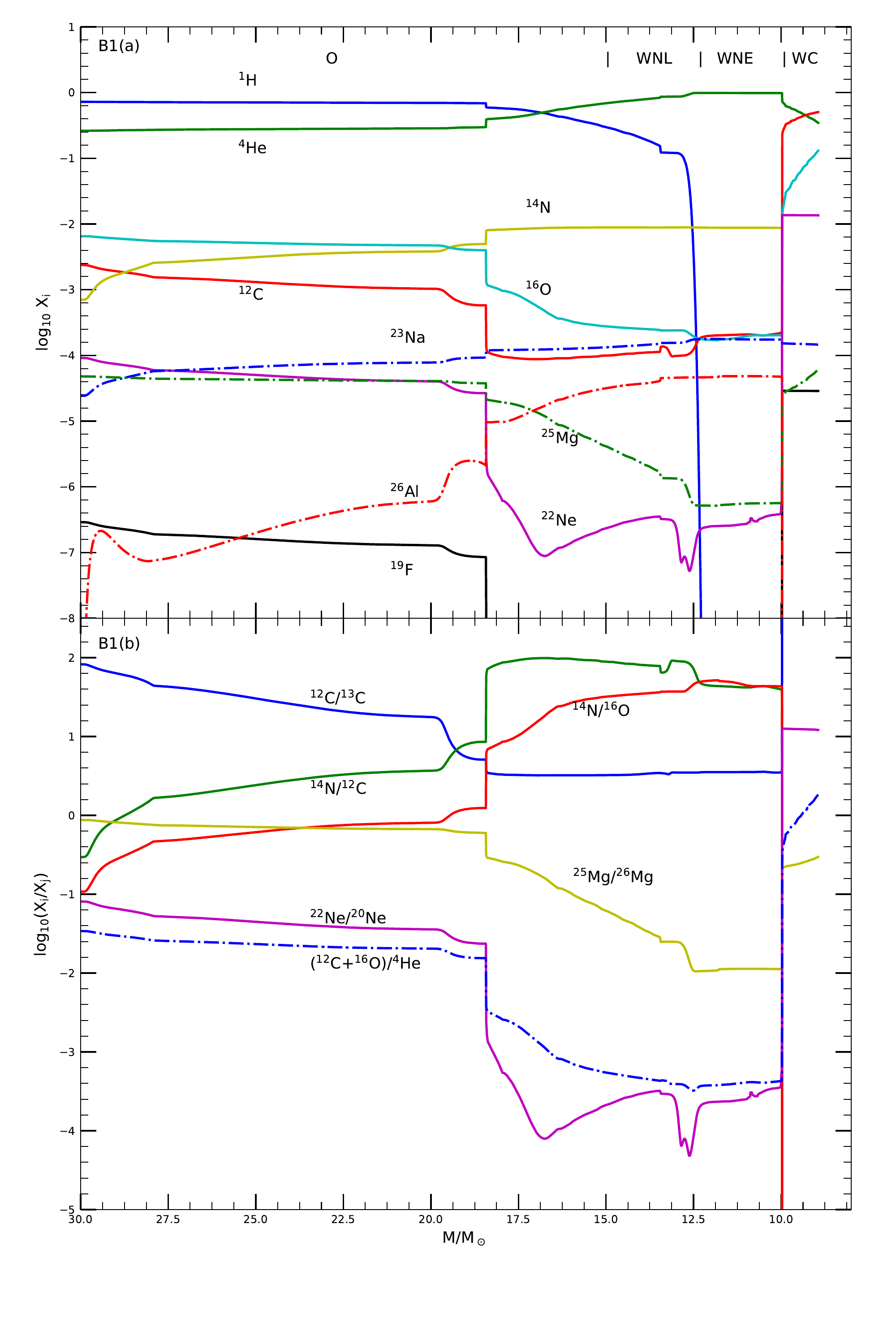}
\includegraphics[width=0.35\textwidth]{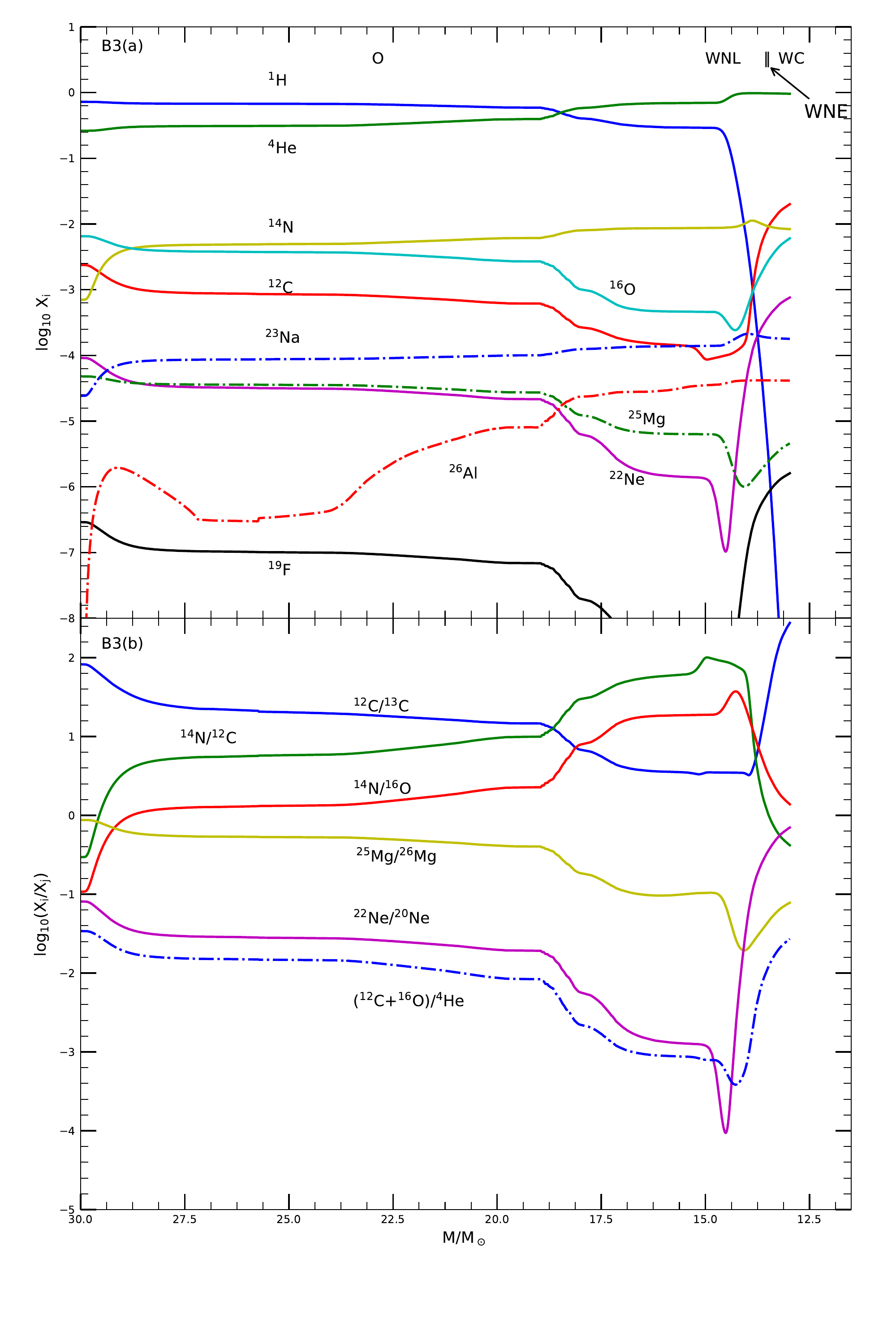}
\caption{Panel S1(a): Evolution of the surface abundances in mass fraction as a function of the actual mass of the star for the nonrotating single star S1 with 30 $M_{\odot}$. Different evolutionary phases are indicated in the upper part of the figure. Panel S1(b): Evolution of abundance ratios in number as a function of the actual mass of the star S1. Panels S2(a) and S2(b): Same as Panels S1(a) and S1(b) but for the rotating model S2. Panels B1(a) and B1(b): Same as the Panels S1(a) and S1(b) but for the primary star in model B1 with initial $\rm P_{orb}=6.0$ days. Panels B3(a) and B3(b): Same as the Panels S1(a) and S1(b) for the primary star in model B3 with initial $\rm P_{orb}=1000.0$ days.}
\end{figure*}

\subsection{Evolution on the Hertzsprung-Russel diagram}
\subsubsection{Single stars}

The WC stage is shown as a green line while the WO stage is indicated as a cyan line in Fig. 1. Various symbols are used to label the beginning and end of the core burning phase: hexagons correspond to the end of the main sequence (MS); pentagrams and tetragonals correspond to the start of hydrogen-shell burning and the end of the core helium burning phase, and triangles and crosses denote the start and end of carbon burning, respectively. The start and end of the mass-transfer phases are marked with numbers; 1- the beginning of the first episode of RLOF, $\rm 1^{'}$ (or 1e): the end of the first episode of RLOF, 2- the beginning of the second episode of RLOF, $\rm 2^{'}$: the end of the second episode of RLOF, 3- the beginning of the third episode of RLOF, $\rm 3^{'}$: the end of the third episode of RLOF.

Panel (a) of Fig. 1 shows the evolution of the single-star models in the Hertzsprung-Russel (HR) diagram. The massive star may lose a large fraction of its initial mass during core hydrogen burning. As hydrogen fuel is exhausted, outward pressure in the core drops, and gravity compresses the star.
Rising heat in the contracting core creates pressure that causes outer layers to expand. The remaining hydrogen burns faster in the shell around the core and generates extra energy, disrupting hydrostatic equilibrium and causing the outer region to expand and cool. This is the mirror effect. During this post-main sequence phase, the outer layers of the star expand to many times their original size on a thermal timescale while the central core contracts. The star shifts from a blue supergiant to a red supergiant with a low temperature. The transition from a blue supergiant to a red supergiant is mainly determined by the combined effects of the hydrogen-rich envelope and hydrogen shell-burning source. High-efficiency hydrogen-shell burning boosts the envelope to expand rapidly. Podsiadlowski
et al. (1992) presented results showing that an important factor in determining the time of transition to
the red giant phase is the fractional core mass $\rm \xi_{c}= m_{c}/M$,
where $m_{c}$ is the mass of the hydrogen-exhausted core and $M$ is
the total mass of the star. A reduction in envelope
mass through stellar winds and a
consequent increase in $\rm \xi_{c}$ favors redward motion in the HR
diagram (Maeder 1984).

The hydrogen-burning shell moves outward in the envelope. If the hydrogen-rich envelope is thick enough, the star can maintain a red supergiant structure. When
the mass of the hydrogen-rich envelope falls below a certain threshold value due to stellar wind, the star turns into a blue supergiant configuration again.  After that, the star turns into a WR object.
The low-mass star S3 goes through a red supergiant stage
without entering the WR regime. The star ignites central helium as a B-type star at the end of the main sequence (MS) and moves blueward on a nuclear timescale.
The 20 M$_\odot$ model reaches $\rm \log (T_{\rm eff}/K )\sim 3.6$ at the red supergiant stage; subsequently, because of the
strong wind mass loss, it evolves back to the blue part of the HR diagram. The star eventually reaches the blue supergiant phase when the central carbon abundance is exhausted.

van Loon et al. (2005) discovered that the mass-loss rate for stars covered with dust during the red supergiant stage is significantly higher than those of visually bright stars. High mass-loss rates can result in stars evolving back from the red supergiant stage to the blue side of the HR diagram instead of exploding as a type II-P supernova. However, Meynet et al. (2015) revealed that models computed with a strong mass loss during the red supergiant phase are still far from becoming WR stars. Higher mass-loss rates during the red supergiant stage simply reduce the time that the star spends as a red supergiant, but do not increase the total amount of mass lost during this stage considerably. The 20 M$_\odot$ model star is predicted to explode before all surface hydrogen is removed by stellar winds, and does not attain the hydrogen-free WNE or WC stages.

In the single-star evolutionary models for masses above $\sim$ 25-30 M$_\odot$, the mass-loss due to strong stellar winds during the MS stage (plus additional mass loss during an LBV stage) strip off the H-rich outer layers, revealing the H-burning products $\rm ^{4}He$ and $\rm ^{14}N$. Such a star is identified as a WN-type WR star. The nonrotating 30 M$_\odot$ star attains the stage of WNL at the age of 6.684 Myr while the corresponding rotating model S2 reaches the stage of WNL at an age of 7.135 Myr. This indicates that the larger age is simply because rotation increases the duration of the main sequence. Rotating models can attain the WR stage earlier in accordance with the mass fraction of He in the core. For example, the model S2 can attain the stage of WC at an age of 7.244 Myr while the nonrotating model S1 does not attain the stage of WC at the end of evolution. These results indicate that $\rm ^{12}C$ and $\rm ^{16}O$, which are produced by the helium burning shell, can be exposed early on by rotational mixing.
Therefore, rotation decreases the minimum mass that is required to attain the WC stage.
The model S4 with a mass of 40 $M_{\odot}$ can completely lose its hydrogen, becoming a WC and finally a WO star. The two observed WO stars nicely agree with the evolution of the rotating
tracks in model S5 with a mass of 35 $M_{\odot}$. Our results indicate that WO stars have evolved
from early-type WC stars with similar surface abundances during their bluewards evolution.
These stars are core-helium burning or post core-helium burning and represent a final stage in massive
star evolution. They
will explode as type-Ic supernovae within a
few thousand years. These WO stars originate from
single rotating stars with initial masses of greater than about $30 M_{\odot}$.

Single-star models with rotation and an initial mass of 30 $M_{\odot}$ produce WC stars with luminoisties of at least $10^{5.62} L_{\odot}$.
The low mass limit for producing a WC star with luminosities above $10^{5.4} L_{\odot}$ from single stars with $\rm v_{ini}=300 km/s$ is about 25 $M_{\odot}$ at solar metallicity (Limongi \& Chieffi 2018).
The single-star evolutionary models with masses above $\sim$ 25 M$_\odot$ would produce WC stars with luminosities above $10^{5.4} L_{\odot}$, which is above the values observed for many WC stars.
However, the enhanced mass-loss rates (compared to the de Jager rates) during RSG and WR stages proposed by Vanbeveren \& Mennekens (2015) and the Potsdam group can cause the star to approach the observational range of low-luminosity WC stars in model S6 (cf. panel b in Fig. 1). The hydrogen envelope in model S6 decreases faster than the one in model S2 from the RSG stage. For example, the mass for model S6 is 8.924 M$_\odot$ at the core helium exhaustion while it is 13.705 M$_\odot$ for the model S2. Although there is a significant temperature discrepancy, the star in model S6 attains the lowest luminosity of $\sim 10^{5.4} L_{\odot}$ because of the reduction of the convective core and then turns into a WO star in its subsequent evolution (Panel (a) of Fig.2.). Strong mass loss during the RSG phase favors a bluewards evolution and lower luminosity WC stars are easier to form according to the formalism proposed by the Potsdam group (Vanbeveren et al. 2007, 2020). This indicates that the envelope stripped by strong winds is still too thick to evolve into the observed range of the low luminosity WC stars. However, the enhanced mass-loss rate can fit the observations of low-luminosity WO stars well and this is therefore a likely channel of formation for these stars.

\subsubsection{The primary star in the HR diagram}

1. Case A mass transfer

From panel (a) of Fig. 1, we can anticipate that binary evolution may play an important role in forming lower luminosity WR stars. In panel (b) of Fig. 1, the evolutionary track of the primary in the close binary system composed of a 30 $M_{\odot}$ star and a 27 $M_{\odot}$ star is also shown for the case where the initial orbital period is equal to 6.0 days. The primary overfills its Roche lobe at point 1 for the first time during the MS phase, until that point, the star lost little mass due to line-driven stellar winds. There are three episodes of mass transfer due to RLOF. We note that strong mass-loss rates occur in two main events. The first event, characterized by a strong decrease in luminosity, takes place between points 1 and $1'$ (loss of 9.506 $M_{\odot}$ by the primary star in a time interval of about 0.415 Myr).
The second event occurs between points 2 and $2'$ (loss of about 0.41 $M_{\odot}$ in a time interval of about 0.161 Myr).
During these mass-transfer events, the mass-loss rates due to RLOF can be up to nearly 100 times stronger than the wind mass-loss rates of O-type or B-type stars whose  characteristic value is about $\rm 10^{-7} M_{\odot}/yr$. Between these two mass-transfer episodes, the star contracts and then expands again due to thermal adjustment. This is because the timescale for losing mass is shorter than the thermal timescale. The core cannot give rise to enough nuclear energy to keep pace with
the expansion of the envelope.

After the core helium burning is ignited, the third episode of mass transfer occurs with the envelope expansion. At the beginning of this last mass transfer, a H-rich envelope of 7.04 $M_{\odot}$ is still present. The star is a WNL star. During the third event of RLOF, the primary star transfers about 3.641 $M_{\odot}$ of envelope to the companion star.
After the third event of RLOF, the primary maintains a hydrogen envelope of 1.3 $M_{\odot}$. As a consequence of three episodes of very large mass loss, the convective core is significantly smaller during the helium burning phase than near the end of the MS. Subsequently, strong mass loss by WR winds reduces the hydrogen-rich envelope by about 4.9 $M_{\odot}$, and the star evolves in the blue region of the HR diagram. The hydrogen burning shell is extinguished and removed readily due to these strong WR winds.
The luminosity decreases again significantly from the end of the third episode of mass transfer (i.e., point $3'$)  to the end of central helium burning. The
radius decreases from 25.12 $R_{\odot}$ to 0.41  $R_{\odot}$. Before the primary reaches the minimum luminosity, it enters the WNE phase.
After that, the helium-burning shell is revealed gradually. The star can turn into a WC star. The lowest luminsities reached during the WC phase are around $\rm Log L/L_{\odot} =5.27$, which is lower than for single-star models but still higher than the  observed lowest luminous WC stars.

With the exhaustion of central helium, the slow contraction of the star induces a higher effective temperature and an increase in luminosity. The primary star does not experience expansion after central helium burning because the envelope above the helium burning shell has been removed by WR winds.
Their cores become more compact during the late evolutionary
stages. Therefore, some early-type WC and WO stars do not display any inflation and agree well with the model predictions in the HR diagram. As the helium shell burns, the shell burning region changes from radiation into convection as a result of the increase in opacity.

2.Case B mass transfer

In the model B2, with an initial orbital period of 20 days, there is a unique mass-transfer episode when the primary crosses the Hertzsprung gap.
The matter is transferred from the primary star to the companion star. This phase lasts about 0.008 Myr. The primary star lost about 8.01 $M_{\odot}$ during the entire mass-transfer phase. The mass transfer is so rapid that the primary is out of thermal equilibrium. The core cannot generate enough nuclear energy to keep pace with the expansion of the envelope. Therefore, the luminosity of the primary falls dramatically and attains the minimum luminosity. When the masses of the two components are equal, the orbital separation reaches its minimum value. From then on, the orbital separation increases when mass is transferred to the companion star, and so the mass-transfer rate of RLOF reduces. The primary star recovers thermal equilibrium again and its luminosity rises.

We notice that RLOF can inhibit the expansion of the hydrogen envelope and hinder the formation of red supergiant stars. This is mainly because little hydrogen is left in the primary star and the primary evolves quickly into a hot and compact helium star. This indicates that the thickness of the hydrogen envelope above the hydrogen burning shell has an important impact on the stellar expansion.
The maximum size of the convective core during the helium burning phase is approximately equal to that near the end of the MS and its mass can exceed half of the mass of the star (cf., Fig. 2). After undergoing different changes in color and luminosity, the star ends its evolution near the He-ZAMS lines. The primary star in model B2 can give rise to an early WC-type star with low luminosity but fails to reproduce the effective temperatures for the observed late WC subtypes with low luminosity, which might be due to the relatively thin helium envelope above the CO core. This also disfavors the occurrence of inflated envelopes. In addition, the effective temperature obtained from observations only reaches temperature in the psudo-photosphere, which is lower than the actual surface effective temperature of the star. When the optical thickness of the wind is accounted for, the temperature may shift to a redder value by about 0.5 dex (Ekstr\"om et al. 2012).

3.Case C mass transfer

In model B3, with an initial period of 1000 days, the binary system undergoes a unique mass transfer 0.013 Myr after the primary star ignites its central helium, which corresponds to
Case C mass transfer. The typical burning temperature for helium burning is around  $2-3\times 10^{8}$ K. Usually the core helium burning is also paired with a hydrogen-burning shell just outside the core.
Central helium burning in massive stars lasts about $10\%$ of the whole lifetime of the star. The luminosity during this phase is comparable to that during hydrogen burning\footnote{It could become lower if the star experiences significant mass loss.}, within a factor of two, but the amount of energy released by helium burning is only about 10 $\%$ of that
released by hydrogen burning. A significant fraction of the luminosity of the star during this phase is also supplied by the hydrogen-burning shell.

The beginning of the mass transfer gives rise to a slight decrease in luminosity, in contrast to Case A and Case B. Contrary to the other systems described in this paper, the primary star in the model B3 spends a large part of its post MS evolution in the red side of the HR diagram, with $\log(T_{\rm eff})<3.9$. After the mass-loss episode, the star evolves towards high effective temperatures and decreases in luminosity, describing a loop in the HR diagram.
We also notice that contrary to the other models, just after the beginning of the helium burning phase, the star loses about 1.34 $M_{\odot}$ through stellar winds. This is mainly because of the mass-loss prescription considered after the MS phase when the star evolves redwards. Later, we find a sharp decrease in the mass of the star due to a short mass-loss episode caused by RLOF, which takes place within about $1.00\times 10^{4}$ years. The star does not experience the red supergiant stage, in contrast to the single star with 30 $M_{\odot}$.
During the helium burning phase, even though the star loses 5.855 $M_{\odot}$ through RLOF, the convective core continues to increase in mass.
The rotating model B3 evolves back to the blue in contrast to the single star S4 with 20 $M_{\odot}$, indicating that rotational mixing, and mass loss due to Roche lobe overflow, promote blueward evolution. Towards the end of the evolution, the star is a WC star with a helium-rich envelope of about 0.24 $M_{\odot}$. Its effective temperature and luminosity can attain $\rm \log T_{eff}\simeq 4.75$ and $\rm \log L/L_{\odot}\simeq 5.27$.  Our Case C evolution behaves very similarly to the single-star model but the enhancement of mass loss is also due to mass transfer via RLOF. This model, due to its different mass-loss history compared to the single-star model, satisfactorily
reproduces lower luminous WC stars with lower effective temperatures.

\subsection{The evolution of stellar mass and the convective cores}
\subsubsection{Single stars}
The upper panel of Fig. 2 shows convective cores and stellar mass for nonrotating and rotating single stars as a function
of evolutionary age. Comparing the nonrotating model S1 with the rotating model S2, we
notice that the amount of mass lost is higher for rapidly rotating
stars during the MS. This can be ascribed to the fact that mass loss through stellar winds can be enhanced by the centrifugal force (Langer 1998). Moreover, the rotational models are slightly overluminous with respect to the nonrotating one and the MS lifetime is increased; this favors the enhancement of mass loss via stellar winds.
Furthermore, convective cores appear to be larger in
the star with high rotational velocity because rotational mixing is very
efficient.
Meridional circulations, which are scaled as the square of rotational angular velocity, are responsible for rotational mixing above the convective
core (Maeder \& Meynet 2000; Song et al. 2018). The larger core induced by rotational mixing leads to higher central temperature and lower opacity in the outer envelope. In fact, the size of the convective core is governed by
radiative pressure which is proportional to the quadrature of temperature $\rm T^{4}$. Therefore, the larger the mass of the star, the larger the convective cores. Rotating stars 30 $M_{\odot}$ have a larger convective core than the nonrotating ones. Furthermore, the main consequence of the rotational mixing is the increase in the lifetime of the core hydrogen burning. The main reason is that fresh hydrogen in the outer envelope is transferred into the central core by rotational mixing. This mixing process increases the fuel supply in the stellar core, extending its lifetime for central hydrogen burning.

\subsubsection{The primary star in binaries}
The bottom panel of Fig. 2 shows convective cores and stellar mass for the primary star in rotating binaries as a function
of its evolutionary age. It is found that the convective core of the primary star is reduced in close binaries compared to
the evolution of the convective core mass in the corresponding single star. For example, at the age of 4.0 Myr, the convective core is 14.26 $M_{\odot}$ for model S2 while it is 13.95 $M_{\odot}$ for model B1. This is mainly because tidal
braking can spin the star down when the spin angular velocity of the primary star is
higher than the orbital velocity. According to Zahn (1977), the synchronization timescale is very sensitive to the ratio of stellar radius R to the orbital separation (i.e., $\frac{1}{\tau_{sync}} \propto (\frac{R}{a})^{8.5}$).
Tidal braking spins the primary star down sharply in the binary system B1 with the shortest orbital period because the orbital separation is shortest in contrast to the wide systems B2 and B3 (cf. Fig.3). Therefore, tidal synchronization significantly decreases the efficiency of rotational mixing. This is because, in this model, the angular momentum transport inside the star is very efficient because spin angular momentum of the star is transformed into the orbit by the tidal braking (Song et al. 2016). For massive stars, the most important contribution to rotational mixing is meridional (Eddington-Sweet) circulation and its the efficiency in mixing is proportional to spin angular velocity $\Omega$ to the power of two. Furthermore, it is clear that strong tides can reduce the differential rotation between the core and the outer envelope and thus decrease the shear turbulence, which may play an important role in rotational mixing. The synchronization timescale $\tau_{\rm sync}$ increases with the initial orbital period to such an extent that the effect of tidal braking can be neglected when the orbital period is equal to or greater than about 100 days in model B3. Therefore, rotational mixing in these models has a similar effect to that in the single rotating star models.

We also can notice that convective cores drop from 11.45 $M_{\odot}$ to 10.13 $M_{\odot}$ during the first episode of RLOF. This result shows that a primary star that loses its hydrogen envelop via RLOF will develop a smaller convective core compared to a single counterpart S2. The removal of mass leads on an immediate drop in pressure throughout the interior of the star, as the weight
of that mass is removed at the stellar surface. This pressure drop is transmitted through the star at the speed of sound, i.e., on a dynamical timescale, resulting in a decreasing of central temperature. Therefore, mass loss tends to quench local nuclear burning, which is extremely temperature sensitive, and the size of convective core diminishes. The star becomes less luminous as a result of the reduction of its convective core (Webbink 2006; Song et al. 2020).

The convective core decreases slightly from 10.09 $M_{\odot}$ to 9.868 $M_{\odot}$
during the second event of RLOF in model B1. Therefore, mass transfer has a slight impact on the size of
the convective core because only a tiny fraction of the hydrogen envelope is removed. The response of the convective core to mass losses induced by mass transfer is different after the MS phase. During the RLOF
the convective core increases from 9.32 $M_{\odot}$ to 9.64 $M_{\odot}$ in model B2 and from 9.37 $M_{\odot}$ to 9.46 $M_{\odot}$ in model B3. This change of behavior is due to two effects. First, during the core helium burning phase, the convective core mass tends to increase in mass as a function of time. Second, the core is less coupled to the envelope, being separated from it by a much stronger density gradient than during the core H burning phase. Thus, removing part of the mass from the H-rich envelope has a very small impact on the evolution of the core.

\subsection{Evolution of the mass-transfer rate due to RLOF}
Figure 4 displays the mass-transfer rate due to RLOF as a function of evolutionary age in binaries. Until RLOF, all three models lose very little mass due to line-driven stellar winds. In the Case A mass transfer described here, there are three episodes of RLOF.
The first event, characterized by a strong decrease in luminosity, takes place between the ages of 5.96 Myr and 6.386 Myr. The maximum mass-transfer rate can attain a value of about $\rm 5.888 \times 10^{-4} M_{\odot} yr^{-1}$, which approximately corresponds to $\frac{M_{1}}{\tau_{\rm KH}}$  where $M_{1}$  and $\tau_{\rm KH}$ represent the
mass and the Kelvin-Helmoltz timescale of the primary star, respectively. Most of the hydrogen envelope remains radiative at the beginning of RLOF.
As mass is transferred from the more massive primary to the less massive secondary, the Roche lobes of the two components shrink due to the orbital shrinkage, but their radii do not. The shrinking Roche lobe of the primary implies that it must lose mass at a higher rate of $\rm 10^{-4}M_{\odot}/yr$. At this high mass-transfer rate, the primary star is out of thermal equilibrium. When the primary turns into the less massive of the two components, subsequent mass transfer from then on will enlarge the orbital separation. Therefore, mass transfer maintains a lower rate (i.e., $\rm \sim 10^{-6}M_{\odot}/yr$). The luminosity of the primary star decreases accordingly.

When thermal equilibrium is restored, both the luminosity and radius rise (cf, Panel b in Fig. 1).
The second event of RLOF occurs between the ages of 6.453 Myr and 6.614 Myr when the envelope of the primary star inflates again near the terminal of the MS. The maximum mass-transfer rate reaches a value of $\rm 3.944 \times 10^{-6} M_{\odot} yr^{-1}$. The mass transfer proceeds at the nuclear timescale of the primary.
The star becomes brighter when the surface mass fraction of hydrogen decreases.
The third mass-transfer event proceeds from 6.646 Myr to 6.666 Myr and the maximum mass-transfer rate can attain $\rm 2.15 \times 10^{-4} M_{\odot} yr^{-1}$. The primary inflates and the resulting high mass-transfer rate proceeds on the Kelvin-Helmholtz timescale.
The total mass loss is about 13.6 $M_{\odot}$ during the three mass-transfer episodes. The primary star loses most of the hydrogen envelope during the episodes of mass transfer. After the mass-transfer episodes, the star is composed of a helium core of 11.53  $M_{\odot}$ surrounded by a 2.7 $M_{\odot}$ H-rich envelope. The primary star appears as a late-type WNL star.

There is a unique mass-transfer episode during RLOF for models B2 and B3. Mass transfer in B2 occurs between the ages 6.908 and 6.927 Myr, while mass transfer in B3 occurs between
6.967 and 6.971 Myr. The maximum mass-transfer rate can reach a value of about $\rm 4.6 \times 10^{-3} M_{\odot} yr^{-1}$ for B2 while it can reach $\rm  5.5 \times 10^{-3} M_{\odot} yr^{-1}$ for B3. The primary star overfills its Roche lobe at the still more advanced stage when the initial orbital period increases; it has a shorter Kelvin-Helmholtz timescale ($\rm \tau_{\rm KH}$) and may have developed a deep convective envelope. The maximum mass-transfer rate which scales as $\frac{M_{1}}{\tau_{\rm KH}}$ becomes larger. Therefore, the mass-loss rate due to RLOF can be up to nearly 100 times stronger than the WR wind mass-loss rates which has a characteristic value of about $\rm 1.0 \times 10^{-5} M_{\odot}/yr$. The total transferred mass is about $\rm 9.144 M_{\odot}$ for model B2 while it is $\rm 5.855 M_{\odot}$ for the model B3. An initially tighter system results in a deeper stripping of hydrogen envelopes. The binary system with the short orbital period is prone to producing early-type WC stars but extreme stripping of the envelope through RLOF disfavors the envelope expansion.

\subsection{The evolution of helium cores}
\subsubsection{Single stars}
Figure 5 shows the helium core for nonrotating and rotating single stars and the rotating primary star in binaries as a function
of evolutionary age.
The higher the initial rotation velocity, the larger the helium core at the core hydrogen exhaustion.
For example, the nonrotating model S1 has a helium core of 11.673 $M_{\odot}$ at the age of 6.38 Myr while the rotating model S2 has a helium core 12.11 $M_{\odot}$ at the age of 6.73 Myr.
This results in a shorter lifetime of helium burning because the helium burning lifetime scales inversely with the helium core mass. The helium burning lifetime for model S1 is 0.51 Myr, while it is 0.48 Myr for S2.

At the core hydrogen exhaustion, the helium mass is 11.73 $M_{\odot}$ at the age of 6.37 Myr for the nonrotating model S1 while it is 12.02 $M_{\odot}$ at the age of 6.73 Myr for the rotating model S2. This is because the helium core mass -which scales with the size of the hydrogen convective core during the main sequence-can be enlarged by the rotation mixing.
During core helium burning, the helium core goes up substantially for both models S1 and S2. This is mainly because the final size of the helium core is closely related to the progression of the H-burning shell at the base of the H-rich envelope. Hydrogen burning in a shell (often the dominant nuclear
burning source) continuously adds fresh helium to the core and this process leads to a larger helium core.
The maximum helium core mass in the nonrotating model S1 has a value of 13.94 $M_{\odot}$ at the age of 6.88 Myr while it has a value of 14.07 $M_{\odot}$ in the rotating model S2. This indicates that the freshly produced helium in the H-burning shell is about 2.21 $M_{\odot}$ in nonrotating model S1 whereas it is 2.05 $M_{\odot}$ in rotating model S2, which implies that the hydrogen shell burning is less active in rotating stars than in their nonrotating counterparts. The main reason for this is that a fraction of the helium has diffused from the core to the hydrogen envelope because of rotational mixing, while hydrogen diffuses in the opposite direction. This process results in a lower hydrogen mass fraction and a lower energy generation efficiency in H-shell burning.
Rotational mixing also leads to enlargement of the CO core mass upon core helium exhaustion because the CO core mass is not therefore inversely proportional to helium core mass. Because the compactness of a star at the beginning of the collapse scales directly
with the CO core mass, a larger CO core mass tends to give rise to a more compact core and a more expanded envelope.

\subsubsection{The primary star in binaries}

After the main sequence, the helium core mass increases gradually with burning of the hydrogen shell. In the binary system B1, the helium core for the primary star is 10.99 $M_{\odot}$  at the age of 6.99 Myr while it is 12.89 $M_{\odot}$ at the age of 7.47 Myr.
The primary star loses most of its hydrogen-rich envelope through three episodes of RLOF. Hydrogen shell burning can be extinguished earlier in this case; the helium core can no longer grow and
may collapse because of the strong stellar wind expected in
the subsequent Wolf-Rayet phase (e.g., Woosley et al. 1995;
Wellstein et al. 2001; Pols \& Dewi 2002).  Hydrogen shell burning can be extinguished at a late time in model B3 (during the core helium-burning
phase), and therefore, the final mass of the
helium core will often be lower for stars in close binaries than in wide binaries. Similar behavior is also seen in the carbon-oxygen core. For example, the mass of the CO core for the primary star in model B1 is 6.9658 $M_{\odot}$ whereas it is 10.785 $M_{\odot}$ in model B3. Furthermore, the mass of the helium core can be eroded greatly by RLOF in the tight system. The mass of the helium core in model B1 loses 3.612 $M_{\odot}$ after the MS whereas it loses 1.293 $M_{\odot}$ in model B3.

The lack of a hydrogen-burning shell because of RLOF results in a higher C/O ratio at the end of helium
core burning, which affects the strength of subsequent carbon
burning and the final size of the iron core (Brown et al. 2001). We also find that RLOF can reduce the central temperature and increase the lifetime of the core helium burning. For example, the lifetime for core helium burning is 0.54 Myr for B1 and 0.52 Myr for model S2 (cf, Table 1 and Table 2).

\subsection{The evolution of surface chemical compositions}
\subsubsection{Single stars}

Panels S1 (a), S2 (a), S1 (b), and S2 (b) of Fig. 6 show the evolution of surface chemical compositions for nonrotating and rotating single 30 $M_{\odot}$ models. There is no surface enrichment of internal chemical products in the nonrotational model S1 until the first dredge up appears. Because of the motion of the convective dredge up, surface nuclides (i.e., $\rm ^{1}H$, $\rm ^{12}C$, $\rm ^{16}O$, $\rm ^{22}Ne$, and $\rm ^{19}F$) decrease because of dilution. Indeed, a deep convective zone can appear for those stars that evolve in the red. This convective zone will reach above the H-shell burning.
These nuclides can be mixed from the mantle to the deep envelope-where because they had been destroyed by the CNO cycle and pp chains during the MS by the convective motion. At the same time, the nuclides $\rm ^{13}C$ and $\rm ^{14}N$ can be produced by the CNO cycle. The element $\rm ^{23}Na$ can be produced via the reaction $\rm ^{22}Ne(p,\gamma)^{23}Na$. The radioactive isotope $\rm ^{26}Al$ is produced by proton capture on $\rm ^{25}Mg$ before it has time to decay. The isotope $\rm ^{26}Al$  is destroyed in the He core of the WN-type star, and thus its abundance is decreased in the winds of WC and WO stars. Therefore, these nuclides  may exhibit surface enrichments.

Furthermore, surface enrichments in internal chemical products for S1 can also be ascribed to the mass removal of hydrogen envelopes via stellar winds after the MS. Markova et al. (2018) noted that the envelope is stripped in the most luminous supergiants by the strong winds ($\rm \log L/L_{\odot} \geq 5.8$ and $\rm \log \dot{M} [M_{\odot}/\rm year] \geq -5.4$). The stellar mass in model S1 reduces from 26.917 $M_{\odot}$ at the end of hydrogen core burning to 14.105 $M_{\odot}$ at the end of helium core burning. The removed hydrogen envelope can eliminate the convective envelope which spans the mass coordinate from 26.6 $M_{\odot}$ to 16.6 $M_{\odot}$ and may attain the position of hydrogen-burning shell. Therefore, the enhancement of chemical compositions can be ascribed to the fact that the CNO products which are generated when the hydrogen-burning shell is exposed by strong stellar winds at the red supergiant stage. As the surface helium mass fraction increases, the star becomes a nitrogen-rich WR star evolving from an O-type star to a WNL star.

In the presence of rotation, surface $\rm ^{14}N$ and $\rm ^{4}He$ can be enriched by the rotational mixing (Meynet \& Maeder 2000; Maeder et al. 2014; Chieffi \& Limongi 2013; Limongi \& Chieffi 2018; Song et al. 2018). The main effect of rotational mixing is to smooth the internal chemical gradients and to facilitate a more progressive arrival of internal nuclear products at the surface (Georgy et al. 2012).
Furthermore, rotation can also enrich the surface $\rm ^{23}Na$ and $\rm ^{26}Al$, as can be seen comparing the model S1 with the model S2. This means that $\rm ^{23}Na$ is always produced by stars in which H-burning takes place via the NeNa cycle. $\rm ^{26}Al$ can be produced in the core by the MgAl cycle if $T \geq 3.0\times 10^{7} K$. Because $\rm ^{19}F$ is destroyed by the CNO cycle, surface $\rm ^{19}F$ element drops rapidly in the rotating star during the main sequence.  However, a portion of $\rm ^{19}F$ can also be made during helium burning by the nuclear reaction $\rm ^{15}N(\alpha,\gamma)^{19}F$ with $\rm ^{15}N$ from $\rm ^{18}O(p,\alpha)^{15}N$ and protons from $\rm ^{14}N(n,p)^{14}C$ (Meynet \& Arnould 1993, 2000; Woosley et al. 2002). We note that, after the central helium burning, the nuclide $\rm ^{19}F$ on the surface is more rapidly enriched in the nonrotating star S1 than in the rotating star S2.

Maeder (2009) presented results suggesting that the behavior of the surface excess of nitrogen is a multivariate function (i.e., stellar mass, evolutionary age, projected rotational velocity, metallicity) for a single rotating star.
As expected, we find that nitrogen enrichment increases with increasing of initial velocity and evolutionary age during the MS (cf, Table 3).
During core helium burning, the nitrogen enrichment factor for the rotating 30 $M_{\odot}$ star increases from 5.22 to 19.95. There are three main reasons for nitrogen enhancement. First, strong stellar winds can remove the hydrogen envelope and expose the hydrogen-burning shell which is richer in nitrogen. Second,
rapid expansion results in larger differential rotation which can strengthen the shear instability. Thus the angular momentum transportation from the core to the envelope becomes more efficient, meaning that outer layers can attain the high rotational velocity which favors efficient rotational mixing and mass removal. For example, we find that the equatorial velocity of the model S2 can attain $\rm 248.04 Km/s$ at the end of central helium burning (cf, Table 2). Third, and most important, the mixing of chemical
elements between the helium convective core and the hydrogen-burning shell, which is induced by rotational
mixing greatly modifies the nitrogen enrichment. As opposed to model S1, where no
shell interactions takes place, models with shell interactions can give rise to more $\rm ^{14}N$. Figure 7 shows the variation of the ratio of surface mass fraction $\rm ^{14}N$ to the sum of the initial CNO elements as a function of the stellar mass for the different models. The maximum ratio for the rotating single star S2 can attain 1.45 before core helium exhaustion. More precisely, fresh $\rm ^{12}C$, which
is synthesized in the central He-burning core is transferred to the hydrogen-burning shell, where it
is quickly converted to the primary $\rm ^{14}N$ and all the other CNO nuclei, whose relative abundances are
dictated by the temperature of the H-shell (Limongi and Chieffi 2018).
This process indicates that the abundances of all
of the nuclei involved in the CNO cycle are increased by this interaction.

However, only a tiny amount of primary nitrogen can be produced in the single rotating model S2 with the solar metallicity.
There are two main reasons for this. First, the stars can lose more spin angular momentum and therefore they
rotate slower because of strong stellar winds. Simultaneously, they are less compact, and therefore differential
rotation and shear mixing are weaker. Second, the H burning shell has a much lower temperature
and therefore resides further away from the core, which disfavors mixing between the two (Meynet \& Maeder
2002).

Furthermore, a very small fraction of $\rm ^{14}N$ that transferred back to the central core is quickly
converted into $\rm ^{22}Ne$ before being converted into $\rm ^{25}Mg$ and $\rm ^{26}Mg$, thereby becoming an efficient primary
neutron source. The freshly synthesized helium which is also brought toward the center can reduce the final $\rm ^{12}C/^{16}O$
ratio in the core because these helium elements favor the conversion of $\rm ^{12}C$  into $\rm ^{16}O$. This process
favors the formation of WO stars.

\subsubsection{The primary star in binaries}

Panels B1(a), B3(a), B1(b), and B3(b) in Fig. 6 show the evolution of surface chemical compositions for the primary star in models B1 and B3.
Surface $\rm ^{4}He$ and $\rm ^{14}N$ in the binary system B1 can attain a higher value at the end of the MS compared to its single-star counterparts S2. This is mainly because surface composition is changed as a result of the outer layers being deeply exposed via RLOF. During three episodes of mass transfer due to RLOF, the factor of nitrogen enrichment in model B1 increases from 5.974 to 12.494. Similarly, the factor of nitrogen enrichment in model B2 increases from 6.700 to 12.55 due to RLOF.
However, the hydrogen-burning shell which is the main energy source can be removed as a result of RLOF in binary models. Before core helium exhaustion, the maximum ratio of the mass fraction of $\rm ^{14}N$ to the sum of CNO can reach 1.17 for model B3 while it is 0.92 for model B2 (cf. Fig.6). Therefore, the production of the primary nitrogen can also be restricted by mass transfer via RLOF in contrast to the situation for single star models.
Moreover, RLOF can drastically reduce the spin angular momentum of the primary star and lower the rotation velocity (cf, Table 2).
For instance, the rotational velocity of the primary in model B1 attains 1.83 km s$^{-1}$ at the end of central helium burning while it is 248.04 km s$^{-1}$ for the single-star counterpart S2.
These facts imply that RLOF does not favor the production of the rotational mixing.

A sharp increase in nitrogen enrichment can be seen in model B3 at the age of 7.406 Myr. This indicates that the hydrogen-burning shell in model B3 is exposed by RLOF during the middle period of core helium burning. We find that the WC stage begins early due to RLOF in system B1 compared to system B2. For instance, the primary star in model B1 attains the WC stage at 7.1064 Myr whereas model B3 reaches the WC stage at 7.456 Myr. The mass fraction of helium in the convective core is 0.068 for system B1 while it is 0.012 for system B3.
In contrast to single stars, the most remarkable feature is that the layers that are rich in $\rm ^{12}C$, $\rm ^{16}O$, $\rm ^{19}F$, $\rm ^{22}Ne$, and $\rm ^{25}Mg$ but deficient in $\rm ^{1}H$, $\rm ^{4}He$, and $\rm ^{14}N$ can be exposed early in binaries. Therefore, the corresponding ratios in number $\rm \frac{^{12}C}{^{16}O}$, $\rm \frac{^{25}Mg}{^{26}Mg}$, and $\rm \frac{^{12}C+^{16}O}{^{4}He}$ increase with the decreasing initial orbital period. The shorter the orbital period, the earlier these layers are revealed. This also indicates that
the minimum mass needed to enter the WC stage decrease with decreasing orbital period because the outer layers can be "peeled off" more effectively in the system with the short orbital period.


\subsection{The profile of various chemical elements at the end of central helium burning}
\subsubsection{Single stars}
Panel (a) of Fig. 8 shows the abundances of various elements as a function of the lagrangian mass in different models at the end of the core He-burning phase.
We find that there is a small amount of hydrogen ($X_{\rm H}=0.183$) at the surface of the nonrotating 30 $M_{\odot}$ star S1 (cf. Table 2) whereas there is no hydrogen at the surface of the rotating 30 $M_{\odot}$ star S2.
There are two main reasons for this. First, rotational mixing in S2 can transfer fresh hydrogen from the envelope to the core and less hydrogen can be maintained in model S2. Second, the mass loss can be enhanced by both the slightly increased luminosity and the longer duration of the MS phase in S2, and therefore the hydrogen-deficient layer is exposed (cf. panel b). Yoon et al. (2010) noticed that the presence of a thin hydrogen layer in some models results in a more extended envelope than in the
corresponding single pure helium star models. The surface convective region in model S2 can be decreased by rotation-enhanced mass loss. The surface abundances in $\rm ^{12}C$ and $\rm ^{16}O$ are enriched in the rotating model S2, compared to the nonrotating model S1 (cf. Table 3). This comes from the fact that rotation leads to an increase in the CO-core mass (see Hirschi et al. 2004). Also, the surface abundance of $\rm ^{22}Ne$ is greatly enhanced in the rotating model S2.

The flat profile of helium in the outer region implies that the development of outer convective zones was restrained in the rotating model S2, compared to the nonrotating model S1. For example, there is an outer convective region which spans from $16.63M_{\odot}$ to $16.48 M_{\odot}$ in the nonrotating model S1 whereas the mass coordinate of the corresponding convective layer is from $16.63M_{\odot}$ to $16.35M_{\odot}$ in the rotating model S2.
This is a consequence of the rotational mixing that proceeded during the evolution. Some freshly synthesized helium diffuses into the envelope, which leads to less hydrogen and smaller opacity. As a result, rotational mixing can restrain the development of the outer convective region.

\subsubsection{The primary star in binaries}
Because the ratio of the central density to the central temperature $\rm \frac{\rho_{c}}{T_{c}^{3}}$ decreases with increasing mass and the near constancy of helium burning temperatures, the central density of the primary star in B3 is lower than that in B1 (cf. Table 2). More specifically, the central hydrogen is used up, and most of the initial metallicity of the star is converted to $\rm ^{14}N$. For solar metallicity, the $\rm ^{14}N$ mass fraction in the helium core will therefore be about $1\%-2\%$; $\rm ^{14}N$ burns by two alpha captures and a $\rm \beta^{+}$ decay to $\rm ^{22}Ne$; the reaction is usually $\rm ^{14}N(\alpha,\gamma)^{18}F(e^{+} \nu_{e})^{18}O(\alpha,\gamma)^{22}Ne$.
In model B1, we noted that the neutron-rich isotope $\rm ^{22}Ne$ is generated during helium burning and the abundance of $\rm ^{22}Ne$ is derived from $\rm ^{14}N$.

At the end of the He-burning phase (see panels (d)), the central abundance of $\rm ^{12}C$ in model B3 is significantly lower than in model B1.
The ratio of carbon to oxygen usually decreases as stellar mass increases. At the same time, the abundances of $\rm ^{20}Ne $ and $\rm ^{24}Mg $ are significantly greater. This is a consequence of helium diffusion into the He core at the end of the He-burning phase. Let us recall that $\rm ^{12}C$ is destroyed by alpha capture (to give rise to $\rm ^{16}O$), while $\rm ^{20}Ne$ and $\rm ^{24}Mg$ can also be produced by alpha capture, respectively.

It is clearly shown in panel(b) of Fig. 6 that mass transfer due to RLOF can effectively reduce the mass of the helium convective core. This can be attributed to the fact that an initially tighter orbit leads to deeper stripping of the hydrogen envelope via RLOF. The central temperature can also be decreased by mass removal. For instance, the central temperature for S2 is $\log T_{\rm c}=8.944$, whereas this temperature is $\log T_{\rm c}=8.937$ for B1 (cf, Table 2). For this reason, the central mass fraction of carbon is higher in model B1 and the corresponding central mass fraction of oxygen is smaller in this model. In general, the higher the $\rm ^{12}C$ mass fraction left by core He burning, the slower is the contraction of the CO core and the shallower the final mass-radius relation of the whole star. The more massive the star, the larger the ratio of oxygen to carbon upon central helium exhaustion. Expansion of the helium envelope during the final evolutionary stages becomes more significant for a more compact carbon-oxygen core (the so-called mirror effect).

\subsection{The evolution of stellar radius}
\subsubsection{Single stars}
Figure 9 shows the variation of stellar radius as a function of central temperature.
After leaving the main sequence, the core of the star contracts and its gravitational potential transforms into internal energy. The central temperature increases rapidly and hydrogen begins to ignite in a shell. The shell-burning source is the main factor responsible for the envelope expansion, through the so-called mirror principle.
The maximum radius can be attained in model S1 and this heavily depends on the fact that the star retains a very thick hydrogen envelope. The variation of the radius is rapid because the stellar radius changes on a much shorter thermal timescale after the MS.

Moreover, this also implies that the star has a larger opacity in the presence of heavy hydrogen envelopes which may favor a larger expansion. At the peak of the first stellar expansion, the energy from hydrogen-shell burning contributes about half of the total nuclear luminosity.
After that the star contracts rapidly. This process can be understood by the fact that the energy contribution from the helium-burning shell increases gradually and outweighs that from the hydrogen-shell burning. The turning point in radius expansion is reached when the stars have approximate energy contributions from two shell sources (Laplace et al. 2020). After that, the helium-shell burning governs the radius expansion because the hydrogen-burning shell has been extinguished due to stellar winds. After central helium exhaustion, the rate of neutrino cooling from the carbon-oxygen core increases and the core undergoes rapid Kelvin-Helmholtz contraction. During this phase, both nonrotational and rotational models display overall expansion. With a sufficient amount of helium in the envelope, this would lead to great expansion of the helium envelope.
The mass of the CO core is somewhat higher in the less stripped model S1 in contrast to model S2. Greater CO mass at core He depletion in model S1 favors stronger contraction due to the mirror effect. This is mainly because the shrinkage of the core can alter the hydrostatic structure of
the star, increasing the density and  temperature at the base of the
hydrogen-shell burning region, which in turn enhances the rate of nuclear
energy generation in the shell. The extra energy produced in the shell
cannot be transported by the envelope, and so it enters the envelope
causing it to cool and expand (according to the virial theorem).
Therefore, the envelope expansion after core helium exhaustion is more prominent in a more compact stellar CO core.

\subsubsection{The primary star in binaries}

The ultra-stripped donor star in B1 is largely composed of helium and, later,
heavier elements. As a result, one may expect the primary star
to be very compact and display small expansions after Roche-lobe overflow. However, it has been
shown that the slightly stripped primary star in B3 can swell and reach great dimensions
in the late stages of evolution. The main reason for this is that the Roche lobe is larger in a wider system and less of helium envelope above the helium-burning shell can be removed by RLOF.
Moreover, a large convective envelope which spans from the mass coordinate from about 2 $M_{\odot}$ to 13 $M_{\odot}$ has formed at the end of evolution. Convective motions will modify the energy transport in the envelope,
which strongly affects the radius of the star.

The convective regions may have typical convective velocities that exceed the sound speed of the
stellar interior. Therefore, convective motions may be highly turbulent and clumped. Gr\"afener et al. (2012) presented results showing
that clumping of the convective flows in the outer parts
of the star could be what seeds the clumping in stellar winds. In stellar models, this clumping effect may increase the opacity
of the material in the subsurface convective layer but can transfer the
average opacity to a higher density. This increase
in opacity, inflates the envelope, leading to a
cooler temperature. The
updated OPAL opacity table has been obtained by Iglesias \& Rogers (1996), which causes a strong iron bump at around $\log_{10} T \approx 5.25$ (cf. fig. 10). The base of the inflated envelope in these models is located around the characteristic Fe-bump temperature. The maximum opacity within the Fe-bump is located at temperatures, ranging from $\rm 5<\log (T/K)<5.5$.

The radius extension for luminous, metal-rich Wolf-Rayet stars was found by Ishii et al. (1999). These inflated envelopes usually contain a strong density
inversion, as is often observed in stellar models (Petrovic et al. 2006). As a sign of inflation caused by the iron peak, the density inversion in the outermost layers disappears when $\log_{10} T > 5.07$.
Moreover, massive helium stars ($> 15 M_{\odot}$ for helium stars) can develop a core-halo structure with a very extended radiative envelope. This structural change is also caused by the large peak of the iron peak opacities, which is more prominent in more massive metal-rich stars. Petrovic et al. (2006) noted that for small values of the stellar wind mass-loss rate, an extended envelope
structure is still present. However, for mass-loss rates above a critical value, for which they derive an expression, Wolf-Rayet radii decrease and
the stellar structure becomes compact. This indicates that, although some WC stars
have inflating envelopes for most of their lifetimes, the inflation
disappears as they lose most of their helium envelopes and
their cores become more compact during late evolution.
Our theory shows that the envelopes of some early-type WC and
WO stars do not inflate, which is due to the absence of helium envelopes and the positions of these stars on the hot side of the helium ZAMS lines. WC stars
of later spectral subtypes have much cooler temperatures in the presence of helium envelopes. The final surface effective temperature tends to be lower for a higher
helium envelope mass that remains until the end, which is in
turn determined by the initial orbital period. The remaining thicker helium envelope facilitates greater expansion.

Moreover, the WR models follow a mass-luminosity relation of approximately
$\rm L_{\ast}\propto M_{\ast}^{1.35}$ (Maeder \& Meynet 1987; Langer 1989). The increase in L/M with mass is the primary driver of the extended envelopes and large radii of massive high-metallicity WC stars, because radiative acceleration is proportional to stellar luminosity and opacity. Therefore, for a fixed  opacity, an evolutionary model with a higher mass of WC stars can attain a higher Eddington factor in its evolution because of its higher L/M ratio (cf. Fig. 10).
It is well known that when the star approaches the Eddington limit, that is, the maximum luminosity they can radiate at, large expansions can be triggered. There are two reasons for the decrease in the luminosity of WC stars in the late stages of evolution. First, the central temperature can be decreased by the previous mass removal due to RLOF. Second,
envelope expansions can give rise to a small temperature gradient inside the star. As the envelope expands, the temperature
throughout the He-burning shell and the density at the top of the He-shell
decrease, in turn decreasing the rate at which nuclear energy is generated.
Therefore, the reduced luminosity in B3 in comparison with the model S2 mainly originates from small temperature gradients.

At the final stage of evolution, helium and carbon shell sources release roughly equivalent luminosities. The layer above the helium-burning shell expands, which results in cooling of the helium-burning shell. Finally, the temperature and density of helium-rich material are too low to sustain helium burning and the helium shell source is extinguished in B1. The maximum envelope expansion is restricted by the orbital period and the mass loss via RLOF. The reduced density due to the expansion has two consequences, a further reduced convective energy transport efficiency, and a reduction in the opacity as the iron peak decreases towards lower densities.

\begin{figure}[h]
\centering
\includegraphics[width=0.50\textwidth]{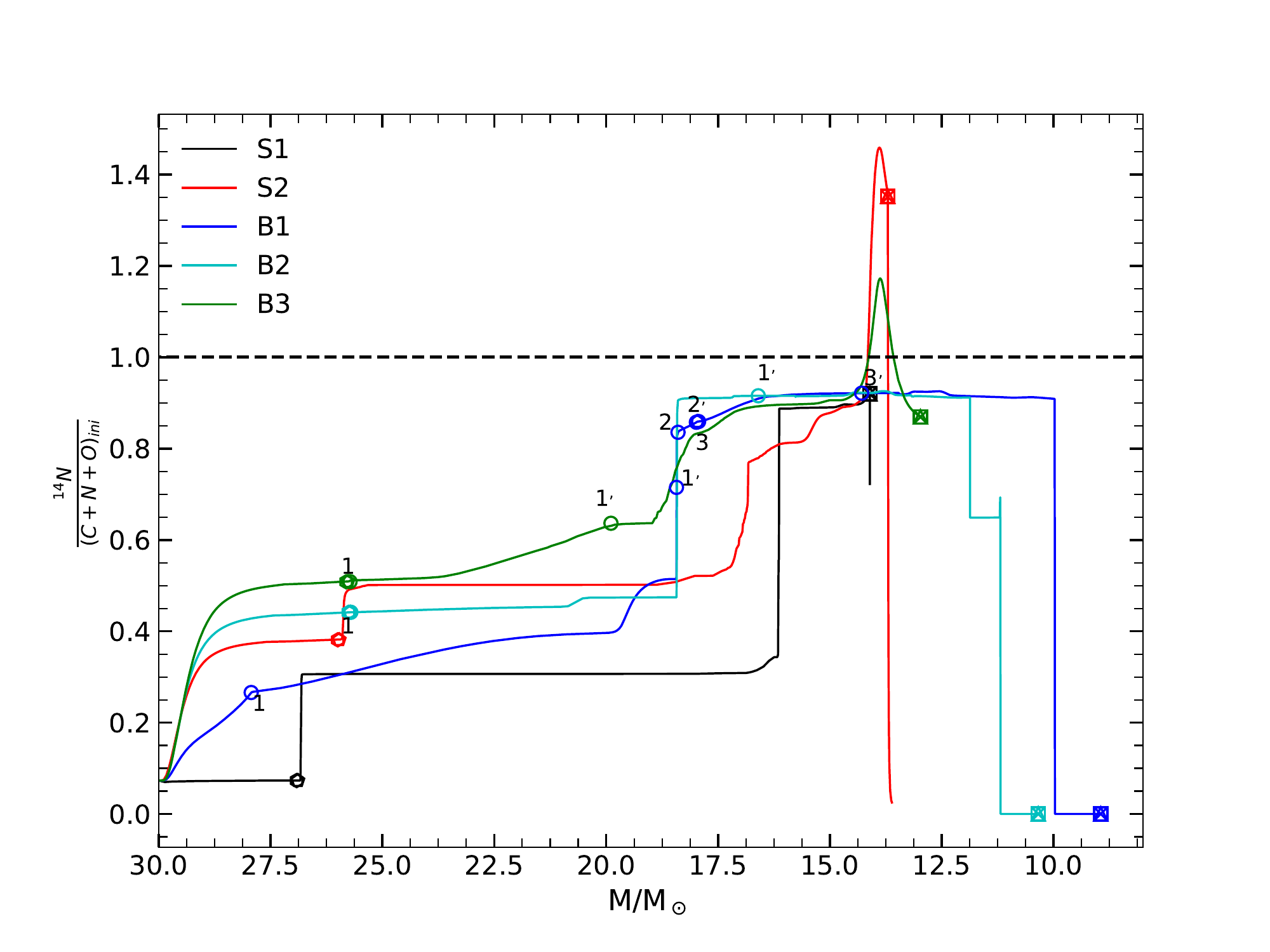}
\caption{Evolution of the ratio of surface mass fraction $\rm ^{14}N$ to the sum of the initial mass fraction of CNO as a function the actual mass of the star in the nonrotating and rotating single stars and the primary stars in rotating binaries.}
\end{figure}

\begin{figure}[h]
\centering
\includegraphics[width=0.50\textwidth]{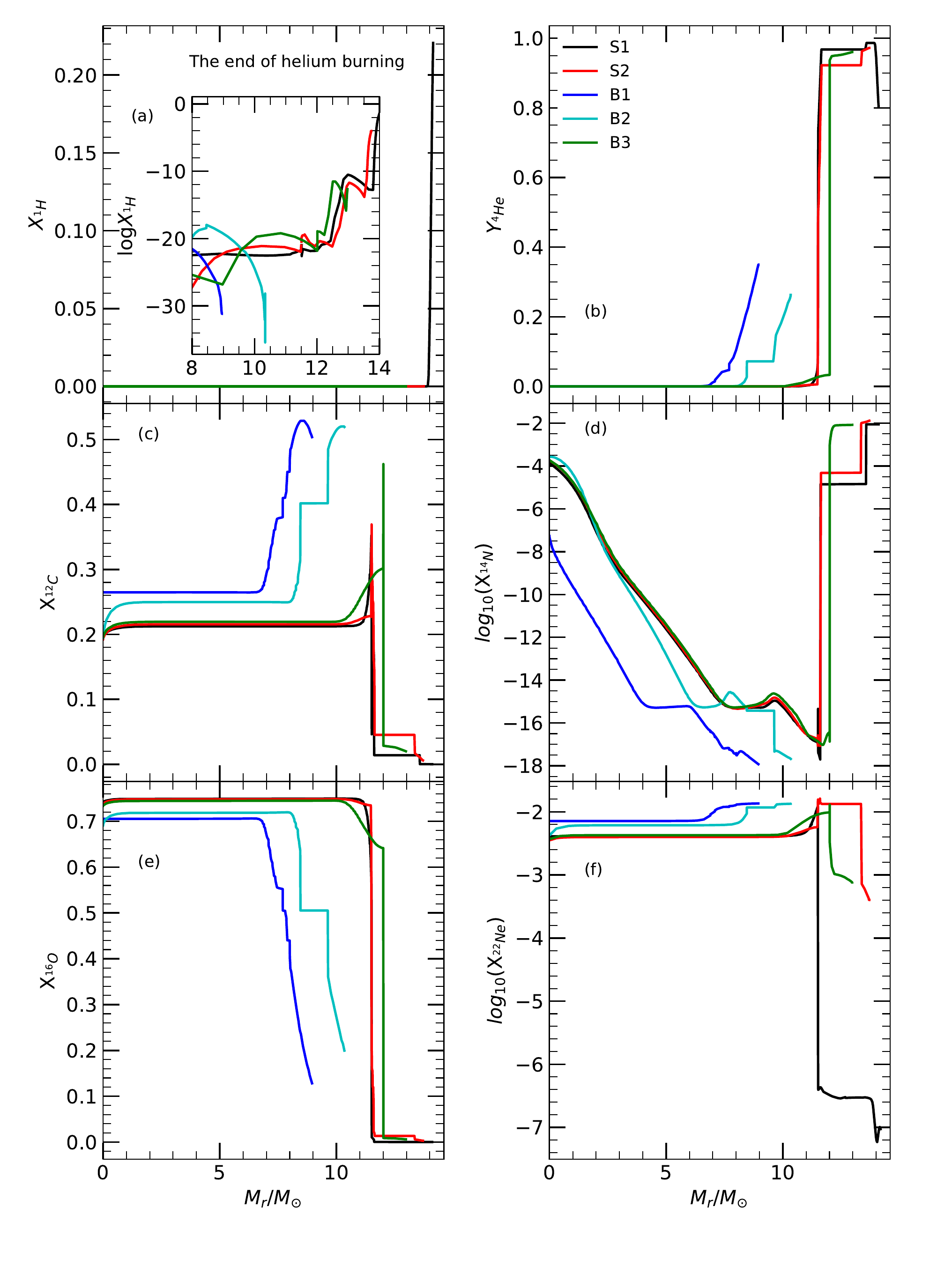}
\caption{Profile of various chemical elements as a function of mass coordinates in single stars with different initial rotational velocities and the donor star in binaries with different initial orbital period at the end of central helium burning.}
\end{figure}

\begin{figure}[h]
\centering
\includegraphics[width=0.50\textwidth]{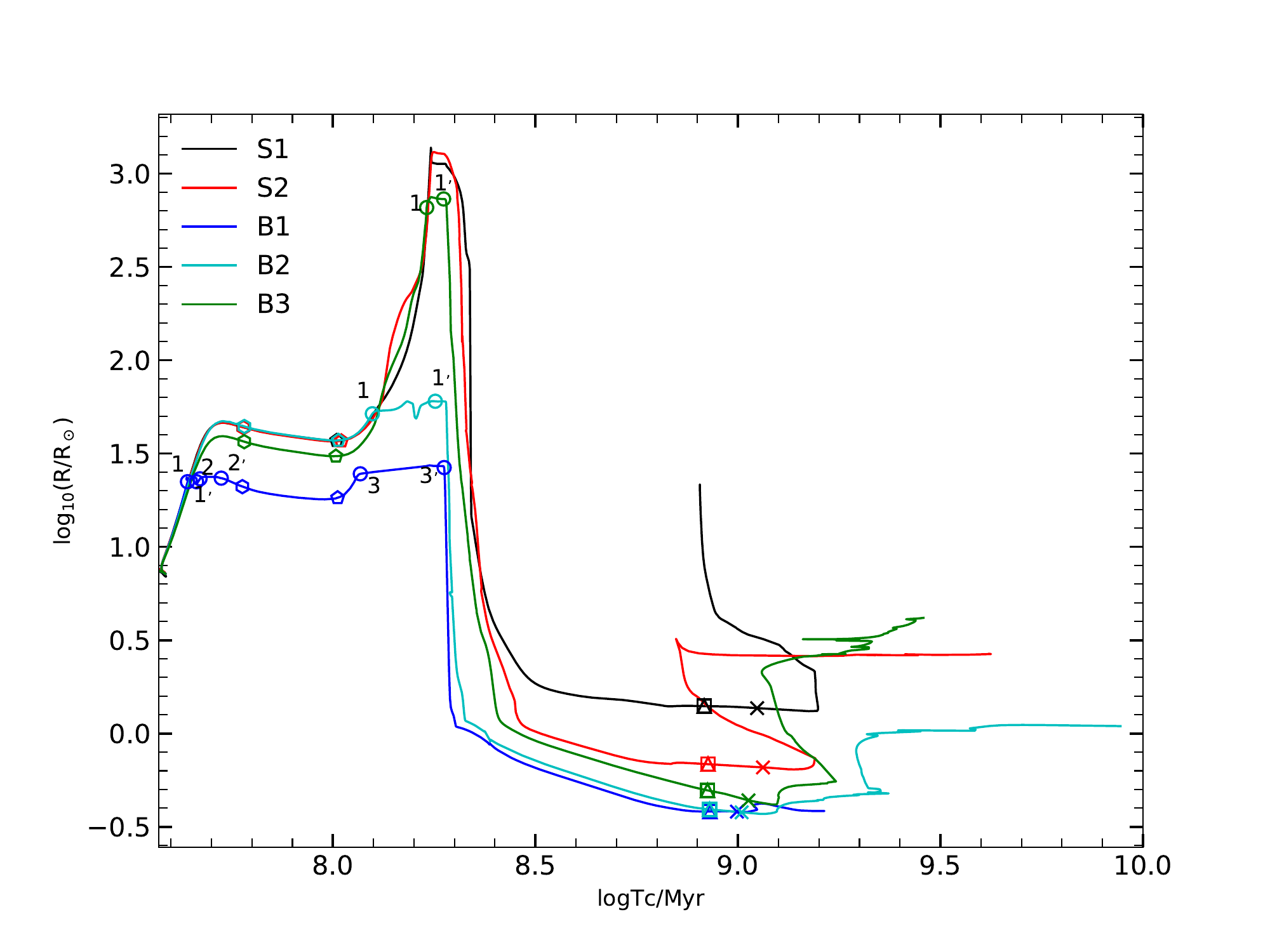}
\caption{Stellar radius as a function of the central temperature of the stellar evolutionary sequences of the models
of different initial velocities in single stars and different initial orbital periods in binaries.}
\end{figure}

\begin{figure}[h]
\centering
\includegraphics[width=0.50\textwidth]{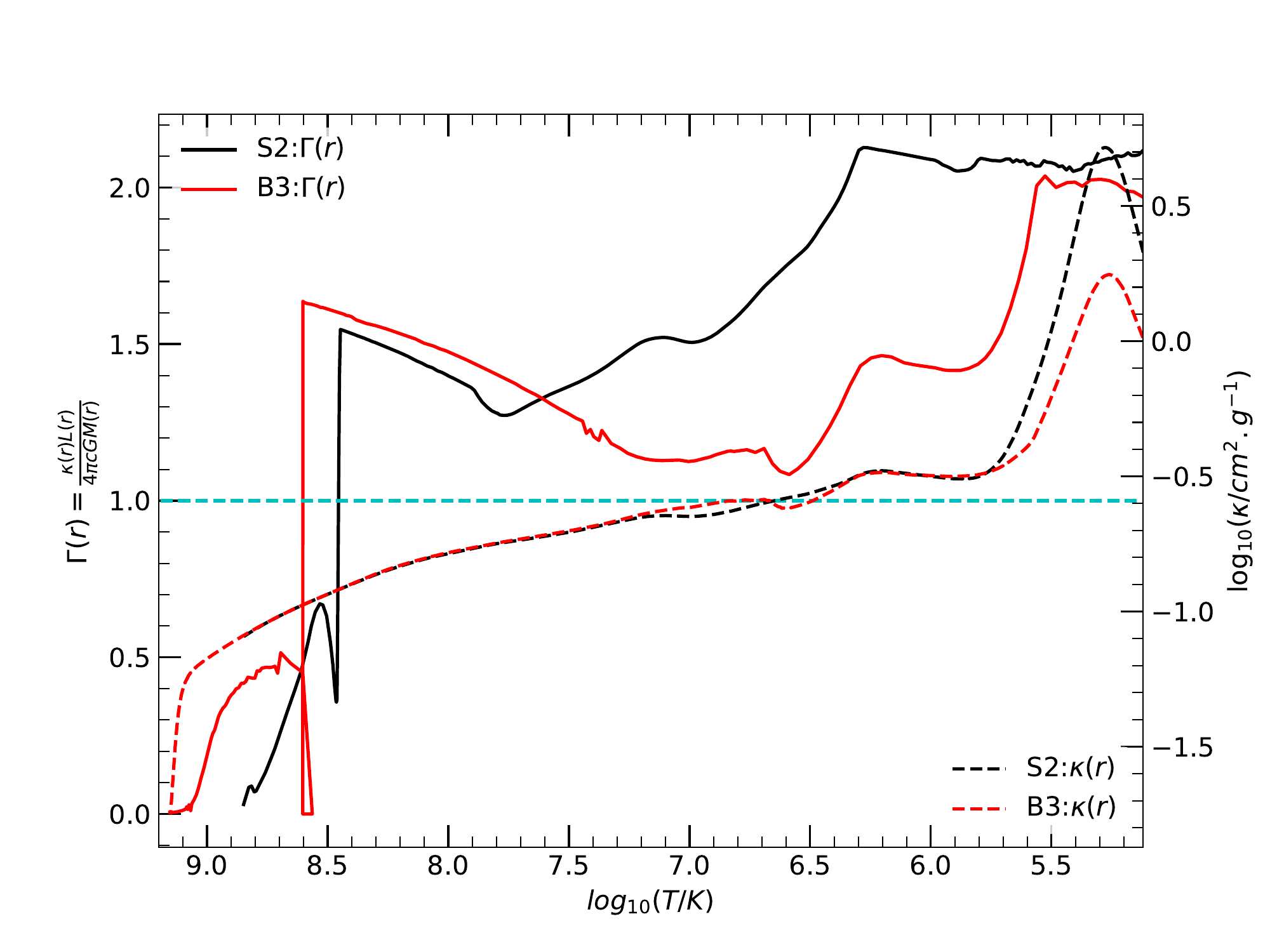}
\caption{Eddington factor and opacity as a function of the temperature inside an initially rotating single 30 $M_{\odot}$ star S2 and a primary star in B3 with the same effective temperature $\rm \log T_{eff}=5.11$ at almost the end of the evolution. The Eddington factor is defined as $\rm \Gamma=\frac{\kappa(r)L(r)}{4\pi cGM(r)}$, where M(r) is the Lagrangian mass coordinate, $\kappa(r)$ is the Rosseland mean opacity, and L(r) is the local luminosity (Langer 1997).}
\end{figure}

\section{Discussion of the validity of the formation of these WC stars in the binary system}

The primary star expands until it overfills its Roche lobe. This channel does not depend on the mass-loss rate of stellar winds but on the orbital period. Therefore, it can work at lower luminosity in contrast to the single-star evolution. The secondary star in the binary system can obtain a fraction of mass through RLOF, and is therefore most likely to become a bright and detectable OB star. About $30\%-40\%$ of the Galactic Wolf-Rayet (WR) stars have a
visible OB-type companion (Vanbeveren \& Conti 1980; Crowther 2007). Shara et al. (2017) found that
12 O-type companion stars in the WR+O binaries rotate super-synchronously and these stars are spun up during the RLOF and mass transfer
of the progenitor binary. This is mainly because mass transfer is generally accompanied by angular momentum transfer
and the mass gainer is expected to spin up.
However, except for a few established binaries, a currently present OB-type companion star is usually observationally excluded (Hamann et al. 2019; Sander et al. 2019). Can the binary channel explain the apparently single WC stars in the observed sample?

We cannot exclude the possibility that these stars are members of long-period binaries. There are four possible physical reasons for the formation of these WC stars in the binary system.
First, the systems that go through Case C mass transfer have long initial orbital periods and the secondary star can accrete a little mass from the primary star.
The companion star is therefore expected to be much fainter in the system with a very small initial mass ratio $\rm q=\frac{M_{2}}{M_{1}}$
than the actual primary star and resides in an orbit that is relatively
wide, which would induce low-amplitude radial velocity
variations of the mass gainers (typically 10 km s$^{-1}$). These
WC stars might be identified as single stars but they have, on the other hand, experienced RLOF. Indeed, a certain fraction of even the
present-day single WC stars might have a binary history, and some
of them might have a yet undetected, low-luminosity companion. Recently, Dsilva et al. (2020) analyzed 12 northern Galactic WC stars and found a clear lack of short-period systems, indicating that a large number of Galactic WC binaries appear to reside in
long-period systems with $\rm P_{orb} > $ 100 days. The analysis of Galactic WC stars performed by Sander et al. (2019) suggests the least massive WC masses are of about 5.8 $M_{\odot}$. This value is much lower than the 10 $M_{\odot}$ obtained by single-star models (Georgy et al. 2012). This supports the view that low-luminosity WC stars could indeed be the result of close-binary evolution.

The model B5 with initial $\rm P_{orb}=500$ days can reproduce the observed region of low-luminosity WC stars in the HR diagram (cf. panel (b) in Fig 1.).
WC stars are believed to be the direct
progenitors of black holes and the orbit does not experience
drastic variation during the short-lived WC phase.
The simulation investigations of the orbital period
distribution of OB+BH binary systems by Langer et al. (2020) revealed that the distribution
of WC binary systems at lower metallicity has a
small peak at short orbital periods $\rm (log P_{orb} \sim 0.7)$ and a larger peak at long
periods $\rm (log P_{orb} \sim 2.2)$ at lower metallicity for the Large Magellanic Cloud. At Galactic metallicity, the corresponding evolutionary tracks
will have larger mass-loss rates and will therefore shift the distribution
towards longer periods. As a rough estimate, the Galactic late WC subtypes with low luminosity can be approximately produced in a wide binary system with an initial orbital period of $\rm P_{orb} \approx 10^{2}-10^{3} days$. A single 30 $M_{\odot}$ star can give rise to a late WC star with high luminosity whereas a single 35 $M_{\odot}$ star can produce WO subtypes. Therefore, the mass range of the primary star that can produce these WC subtypes is approximately estimated to within the range of $30-35 M_{\odot}$.

On the other hand, these WC stars might be runaway stars, kicked off when the primary star exploded in a supernova event. For example, WR136 is a runaway star (Tetzlaff et al. 2011).
As an alternative possibility, avoiding a bright companion star, the evolution of a common envelope has been suggested.
A common envelope phase occurs when the orbital period is on the order of years. If the secondary star is originally
a very low-mass star, it might eject the hydrogen envelope without accreting much mass during common envelope evolution (Kruckow et al. 2016). Common-envelope
evolution gives rise to orbital shrinkage, resulting in a very
short-period binary system that may merge. Therefore, the binary system cannot exist and may lead to an
apparently single WC star. If it survived, the faint companion star with low mass might be very difficult to detect. This means that common envelope evolution (meaning no mass transfer and thus no spin up), which excludes any accretion onto the secondary
star with low mass, may partly be responsible
for the formation of the apparently single WC stars.
Finally, we may consider the possibility that the current WC star was originally the secondary star of a binary system and served as the accretor in the first RLOF. The primary star explodes or collapses to a compact object. If the binary system maintained bound after explosion, reverse RLOF might occur, this time from the original secondary star to the compact object. This process can strip off the hydrogen envelope from the secondary star and turn it into a WC star. This process can lead to a WR and a compact companion system. For example, Moffat et al. (1982) suggested that WR124 might be a binary hosting a compact object.

A particularly challenging object is the WC9 star WR 119.
The least luminous in WC samples, with $\rm \log L/L_{\odot} = 4.7$, and with a current mass of
approximately 6 $M_{\odot}$, this star is most likely the product of the binary evolution,
although there is no clear evidence for a close binary system.
Sander et al. (2019) suggested that this system might be an interesting example of a stripped envelope star originating
from a primary star with $\rm M < 18 M_{\odot}$ that lost its hydrogen envelope
to a companion star and eventually shifted from a WN to a WC stage.

\section{Summary and Conclusion}
Observations of Galactic low-luminosity WC stars indicate these stars are located in regions hotter than log$\rm T_{eff}=4.5$, and their luminosities $\rm \log L/L_{\odot}$ range between 4.9 and 5.4. Previous single-star evolutionary models cannot provide a fully satisfactory explanation for their positions in the HR diagram. Binary evolution is a promising channel, allowing enough mass to be removed to reach the lower luminosity of observable samples. The binary models can span a wider region in both luminosity and temperature than the single-star models. The main results of this work are as follows.

(1)Rotation has two effects on the evolution of massive stars. First, rapid rotation enhances mass loss by reducing the depth of the potential from which mass must escape and therefore increases the chance of forming WR stars during the RSG stage, especially strongly at high velocity. Rapid rotation can cause stars to evolve more quickly away from the red supergiant phase and thus decreases the minimum mass that is required to attain the WC stage.   Therefore, rapid rotation has an impact on the WC/WN ratio by shortening the WN lifetime. Rotation contributes to the mixing of core fusion products through the radiative envelope, increasing surface abundances of heavy elements and stellar winds.

(2)Rotational mixing can increase the He or CO core  mass and decrease the final $\rm ^{12}C/^{16}O$ ratio in the core. A low-mass star with 20 $M_{\odot}$ does not go through a WR stage and such a model ends its evolution as a supergiant. The more massive star (i.e., $M \sim 30 M_{\odot}$ ) can evolve directly from a red supergiant to a WR star in models S1 or S2. The rotating star S2 can reach the late-type WC stage with high luminosity whereas the nonrotating counterpart S1 merely attains the WNL stage at the end of the evolution. Moreover, $\rm ^{12}C$ and $\rm ^{16}O$, which are produced by the helium-burning shell, can be exposed early by rotational mixing and this favors a higher metallicity and a larger opacity. The mass of helium and carbon-oxygen cores is enlarged by rotational mixing.

(3)Before RLOF, we find that nitrogen enrichment is lower in binaries than in single stars. This fact indicates that rotational mixing can be reduced by a lower synchronous rotational velocity.
Nitrogen enrichment is greater for the initial wide system B3 compared to system B1. This can be understood by the fact that tidal braking has a slight impact on the rotational mixing in an initially wide system.
The primary star in binary systems can evolve into a WR star because of stripping by a companion star rather than the inherent mass loss via stellar wind. Extreme stripping of the envelope through RLOF disfavors redward motion in the HR diagram because of the absence of envelope expansions. The primary star that goes through Case A mass transfer can avoid the red supergiant phase and this causes the star to burn most of its helium during a WNL or WNE stage. The primary star ends its evolution as an early-type WC stars with high luminosity and the cores of the binary become more compact during the late evolution.
The primary is expected to be hot and hydrogen depleted. The primary star is also a hot WC star in the system with an initial orbital period of $\rm P_{orb}=20$ days but displays little envelope inflation because of its thin helium envelope.

(4)The primary star responds to the decrease in mass by adapting its internal structure during the MS and thus the size of its convective core reduces. However, the mass of the convective core does not decrease when RLOF occurs after the main sequence. RLOF increases the probability of a star losing its hydrogen envelope and results in fewer RSGs, more WR stars and more Type Ib/c SNe in contrast to single-star models. The donor star is hotter and more compact than its single-star counterpart because this latter keeps more hydrogen in its envelope. As a result, the fraction of WC stars can be produced through the binary channel, and therefore the number of WR stars that can be produced, should be high. The minimum mass for stars entering the WC stage decreases with decreasing initial orbital period.

(5).We find that the later the RLOF occurs, the closer the final state of the primary star is to the observed late-type WC stars with low luminosity in the HR diagram. The late WC stars with low mtallicity have more chance of being produced with an orbital period ranging from 100 days to 1000 days.
In the system with an initial orbital period of $\rm P_{orb}=1000$ days, the primary star can turn into the observed late-type WC star with low luminosity due to significant envelope expansion. Our results are almost is nearly independent of the adopted accretion efficiency factor $1-\beta$.
Heavy helium envelopes can be heated by the helium burning shells and favor envelope expansion. The helium envelope in the wide binary system is not significantly stripped but surface carbon and oxygen are enriched beyond core helium-burning. Envelope inflation tends to be suppressed upon strong mass loss. In the late stages of evolution, the temperature gradient of the primary is smaller than that in the single-star counterpart and thus its luminosity decreases rapidly with the expansion of the envelope.

(6)There are four physical reasons for the occurrence of inflated envelopes. First, less helium envelope can be transferred to the companion star in the wide system. Heavy helium envelopes can be heated by the helium-burning shell and this creates the necessary conditions for envelope expansion. Second, envelope inflation is related to the proximity of the stellar luminosity to its effective Eddington luminosity, that is, to a high luminosity-to-mass ratio or a high envelope opacity. The increase in L/M with mass is the primary cause of the extended envelopes and large radius of massive high-metallicity WR stars. For a fixed opacity,
more massive stars are more prone to inflation due to a higher Eddington factor. Third, the compactness of a star at the beginning of the collapse scales directly
with CO core mass. A larger CO core mass tends to give rise to a more compact core and a greater expansion of envelope due to the mirror effect. RLOF can reduce the mass of both helium and CO cores because hydrogen-burning shells have been removed. Finally, the iron opacity bump also greatly increases the occurrence of inflated envelopes.

\begin{acknowledgements}
 This work was sponsored by the  National Natural Science Foundation
of China (Grant Nos. 11863003, 12173010), Swiss National Science Foundation (project number 200020-172505),  Science and technology plan projects of Guizhou province  (Grant No. [2018]5781).
We are very grateful to an anonymous referee for his/her valuable suggestions and very insightful remarks, which have improved this paper greatly.
\end{acknowledgements}

\begin{appendix} 
\section{The effect of accretion efficiency}

The evolution of the mass-transfer rate from the donor star is presented in Fig. A.1. We investigate three different accretion efficiencies (i.e., $\dot{M}_{2} = (1.0-\beta)\dot{M}_{1}$) in Case A mass transfer.
We find that the greater the mass of the matter that can be accreted by the accretor (i.e., $\rm 1-\beta$), the greater the maximum mass transfer rate is in the first event of RLOF.
However, we notice that the opposite trend appears in the second event of mass-transfer. These differences are related to changes in the
orbital separation (cf. panel a in Fig. A.2.). Moreover, the small amount of hydrogen envelope remaining after the first event of RLOF
is unfavourable for the occurrence of the second event of RLOF.
The rejected matter during RLOF, like stellar winds, always increases both the orbital period and the orbital separation  because the rejected matter from the gainer tends to widen the orbit. Furthermore, a significant difference in the mass-transfer rate evolution as a
function of the accretion efficiency is in the time spent by the star from the end of the
first RLOF to the onset of the second one. A higher accretion efficiency $\rm 1.0-\beta$ during the first event of RLOF can significantly postpone the
onset of the second event of RLOF because the relaxation time which restores thermal equilibrium may be increased by the excessively stripped hydrogen envelope. The total amount of mass transferred by RLOF is almost independent of $\beta$ while the mass of the secondary star strongly depends on this parameter. The luminosity and mass after RLOF are almost unaffected by the matter-transfer efficiency.

In panel (b) in Fig. A.1, we investigate four different accretion efficiencies during Case C mass transfer. We find that the mass-transfer rate in Case C is almost insensitive to the adopted accretion efficiency factor $\rm 1-\beta$. Therefore, the evolution of the orbital period slightly depends on the accretion efficiency. These results indicate that the accretion efficiency has a slight impact on the evolution of the WC star in Case C mass transfer.

\begin{figure}
\includegraphics[width=0.50\textwidth]{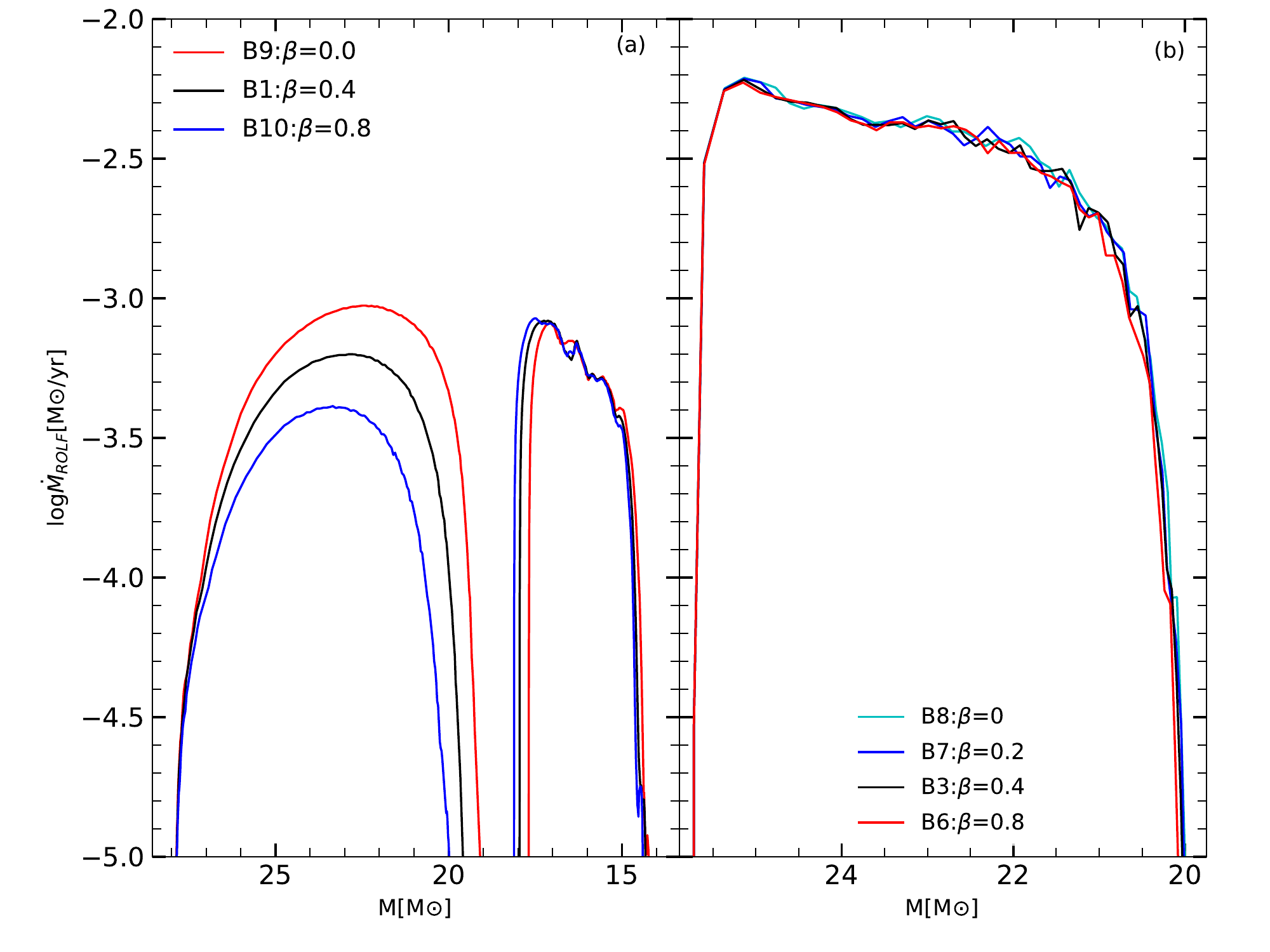}
\caption{Mass-transfer rate in Case A (panel a) and Case C (panel b) from the donor star as a function of the mass of the donor star. In each
panel, we show the different values of the accretion efficiency $\rm 1-\beta$ in Case A and Case C mass transfer in this work.}
\end{figure}
\begin{figure}
\includegraphics[width=0.50\textwidth]{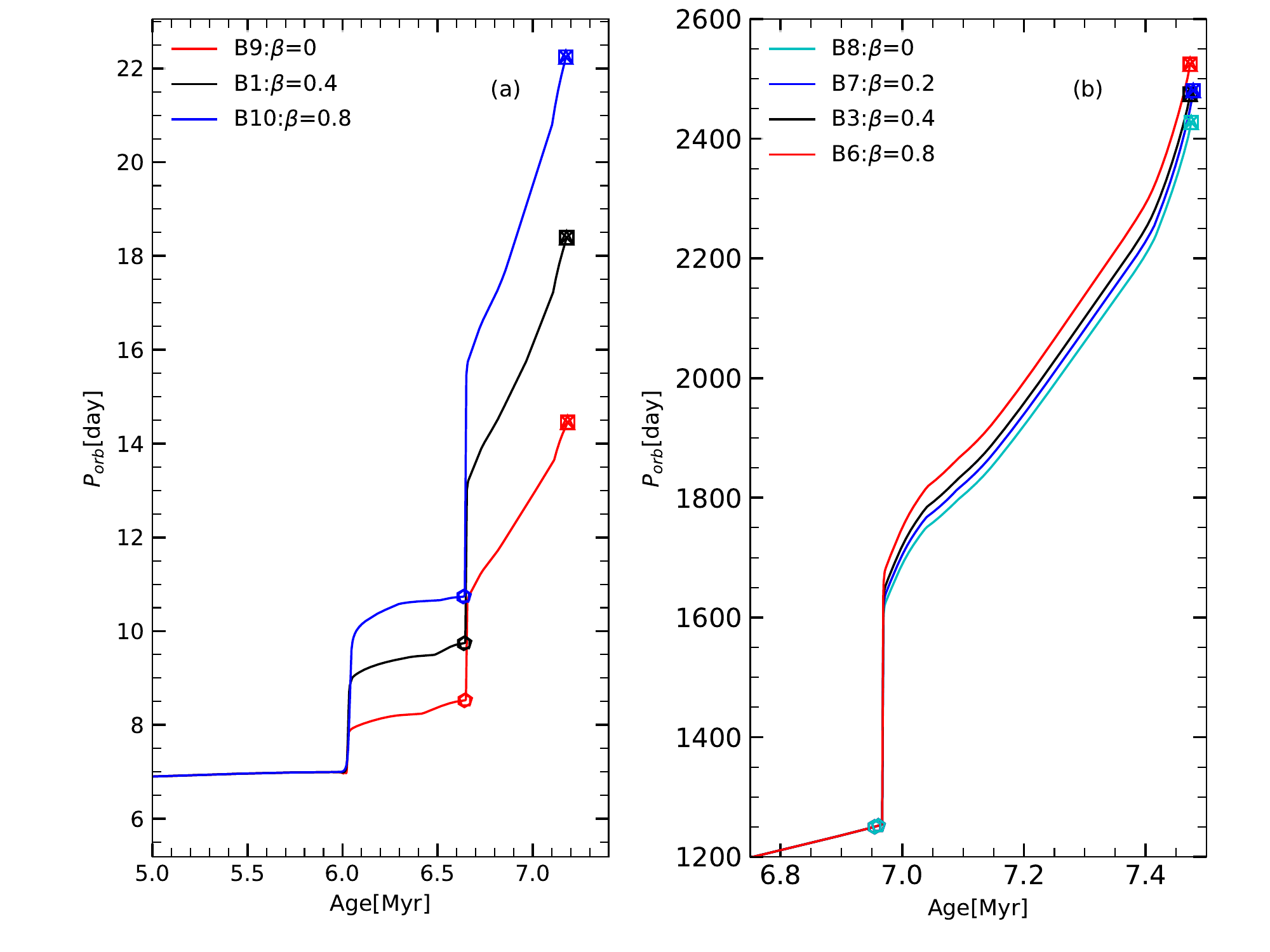}
\caption{Temporal evolution of the orbital period for the different values of the accretion efficiency $\rm 1.0-\beta$
considered in both Case A (panel a) and Case C (panel b) mass transfer.}
\end{figure}

\end{appendix}

\end{document}